**Mechanical mapping of thin elastic films and living cells with spherical tip atomic force microscopy probes at large indentations**

*Gabriel Gomila*[*,1,2], Mauricio Cano[1,2], Beatriz Cantero[1,2], Sophie Strawbridge,[1,2] Lara Aiassa[1], Loris Rizzello,[3,4] Giuseppe Battaglia,[1] Eleni Dalaka[1], Jordi Comelles[1], Annalisa Calò[*1,2]*

[1]*Institut de Bioenginyeria de Catalunya (IBEC), The Barcelona Institute of Science and Technology, c/ Baldiri i Reixac 11-15, 08028, Barcelona, Spain.*
[2]*Departament d'Enginyeria Electrònica i Biomèdica, Universitat de Barcelona, C/ Martí i Franquès 1, 08028, Barcelona, Spain*
[3]*National Institute of Molecular Genetics (INGM), Via Francesco Sforza, 35 20122 Milano, Italy.*
[4]*Department of Pharmaceutical Sciences, University of Milano, Via Luigi Mangiagalli, 25 20133 Milano, Italy.*

*Corresponding Authors: ggomila@ibecbarcelona.eu, acalo@ub.edu

*Abstract*—An analytical model to quantify large indentation force curves acquired on elastic thin films and living cells with spherical tip Atomic Force Microscopy (AFM) probes is presented. The model accounts for the bottom effect in the whole indentation range and overcomes the limitations of Sneddon’s and Hertz's contact models, which are valid for semi-infinite thick samples, and of paraboloid tip models with bottom effect correction (BEC) that are applicable to spherical tips only for relatively small indentations. The model is experimentally validated with force volume measurements on polyacrylamide (PAA) hydrogel thin films, where an excellent agreement is obtained. The accurate correction of the bottom effect demonstrates that the intrinsic Young's modulus of PAA thin films increases for thickness below a critical value (~15 µm). The model also shows excellent agreement with force curves acquired on live macrophages, providing accurate Young's modulus values for these very soft cells ($E$~200 Pa). Young's modulus values extracted with the proposed model significantly differ from those obtained from Sneddon’s or paraboloid models with BEC, whose values deviate by 100% and -25%, respectively.  Results show the potential of the proposed model for analysing force curve measurements with spherical tips at large indentations on thin film elastic materials and living cells at the micro and nanoscale.

## 1. Introduction

Atomic Force Microscopy (AFM) is widely used to measure the micro and nanomechanical properties of soft materials, like thin films, living cells and tissues [1], [2], [3], [4]. Very often, these studies are performed by using spherical tip probes [5], [6], [7], [8], [9] and [10], as they offer a well-defined, simple geometry and a gentle distribution of strains and stresses, helping to preserve the integrity of delicate samples [9]. Spherical tip AFM probes can be fabricated in different ways, such as by gluing a colloidal sphere to a tip-less cantilever [11] or by electron beam deposition of a sphere on a conventional AFM tip [12] (Figures 1a-1b) with radii ranging from tens of nm to tens of µm. For thick samples, static force-indentation curves acquired with spherical tip probes can be quantified with Sneddon's model (or Hertz's model for small indentations) to obtain the local Young's modulus of the material [13], [14], [15].

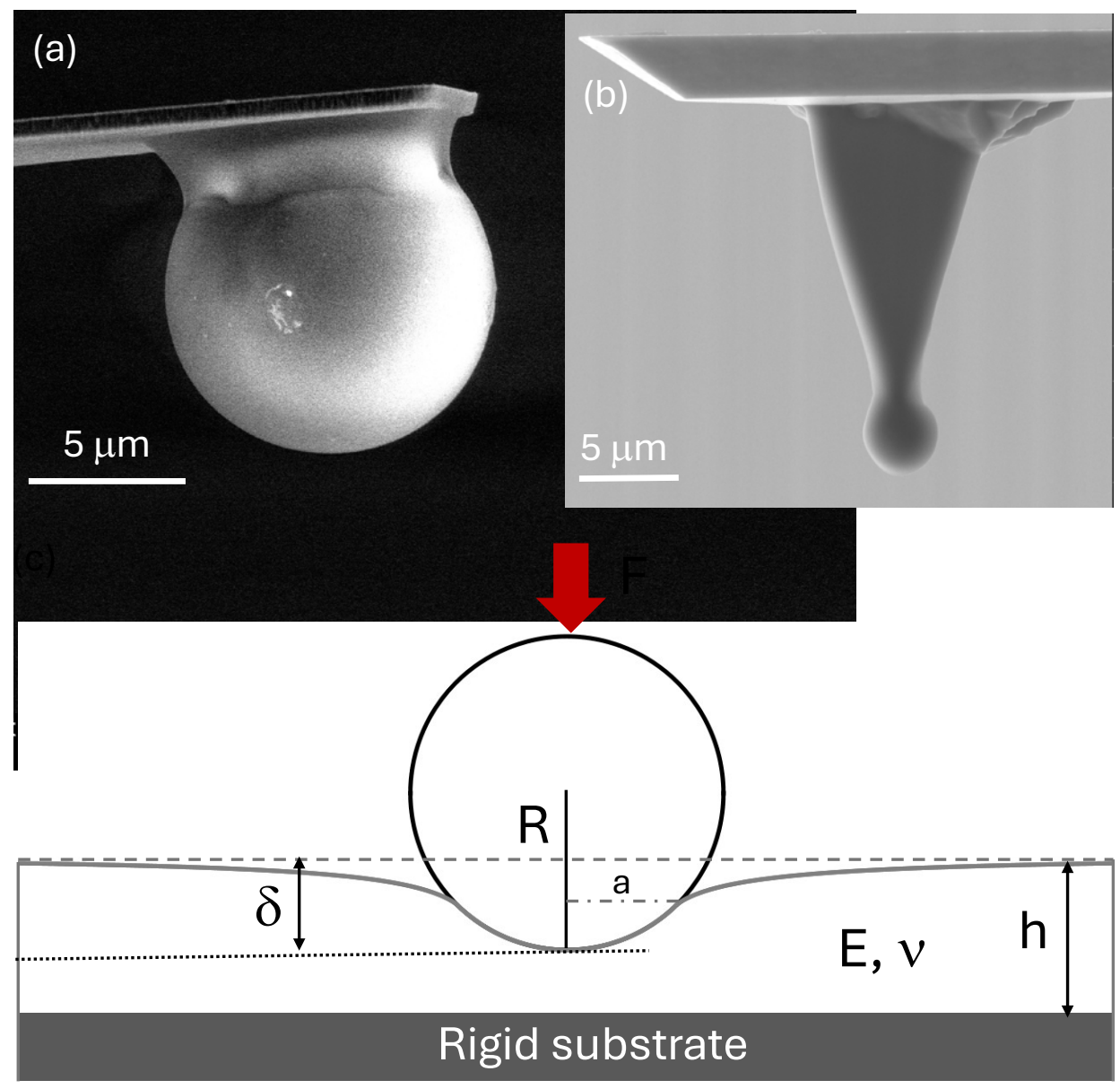

**Figure 1.** (a) and (b) Scanning Electron Microscopy images of colloidal and electron beam deposited spherical tip AFM probes, respectively. (c) Schematic representation of a sphere of radius R indenting an elastic thin film of thickness h, Young's modulus E and Poisson's ratio ν bonded to a rigid substrate under the application of a frictionless loading force F. a is the contact radius and δ the indentation.

For thin samples and spherical probes, the semi-infinite thickness approximation breaks down quickly, as the contact radius *a*, becomes comparable to or exceeds the film thickness [16] (Figure 1c), making force-indentation measurements sensitive to the bottom substrate stiffness [16], [17], [18]. If the quantitative analysis of the measurements performed with models is valid for semi-infinite thick materials, such as Hertz's or Sneddon's models [13], [14], [15], then an apparent Young's modulus is extracted; the values depend on the thickness of the material and on the stiffness of the bottom substrate (e.g., for rigid substrates, materials appear stiffer). To account for the bottom stiffness effect, bottom-effect corrected (BEC) analytical models have been derived for different AFM tip geometries (e.g., flat cylindrical punch, cone, paraboloid, capped cone, etc.) [5], [16], [19], [20], and [21]. However, for spherical probes, an analytical model valid for large indentations does not exist yet. Paraboloid tip models with BEC correction [5], [16] and [20] can be applied to spherical tip probes only for relatively small indentations (typically indentations up to 30-40% of the tip radius). On the other hand, the infinite series expansion parametric analytical solution of Dhaliwal and Rau [22] for a spherical indenter is valid only for contact radius, $a$, smaller than the sample thickness, $h$, i.e., $a/h<1$, leaving many cases of practical interest outside its range of applicability. Spherical phenomenological models based on numerical simulations have been proposed by others [23], [24], and [25]. However, these models cover only restricted ranges of thicknesses, indentations, or both and therefore lack generality.

Finite element numerical simulations could alternatively be used to model spherical probes in situations not covered by existing analytical models [18], [23], [26]; however, their use in practical applications is cumbersome.

To overcome these limitations, we present an explicit analytical model to describe force-indentation curves acquired with spherical tip AFM probes on elastic thin samples valid at large indentations and across the whole range of film thicknesses. The model accounts for the bottom stiffness effect and is valid for laterally infinite thin films made of homogeneous, isotropic, and linear elastic materials bonded to a rigid substrate and for non-adhesive frictionless contacts. The model has been derived using the functional dependency predicted by Dhaliwal and Rau's analytical solution [22] and has been generalised to the range $a/h>1$ using numerically calculated data. The analytical model describes the numerical data with a relative error of less than 1.3% across the whole range of indentations and sample thicknesses, i.e., $\delta_{max}\sim\min(h,R)$. The model is validated experimentally with measurements on PAA hydrogel films of variable thickness, demonstrating its potential to evaluate intrinsic variations in the material's mechanical properties, particularly in situations where the bottom stiffness effect plays an important role. In addition, the model's applicability to living cells at large indentations is shown.

## Results

### 2. Analytical model for a spherical tip AFM probe indenting an elastic thin film on a rigid substrate at large indentations

Figure 2 (black symbols) shows a force-indentation curve numerically calculated for a spherical indenter of radius *R,* indenting a thin elastic film of Young's modulus *E*, Poisson's ratio $\nu=0.5$ and thickness $h=1.25\cdot R$ bonded to a rigid substrate for the case of non-adhesive frictionless contact. The numerical results were obtained by solving the corresponding contact mechanics integral equation according to the formulation in Ref. [27] (see Materials and Methods). Figure 2 also displays, for comparison, the numerically calculated data for a paraboloid tip of radius of curvature *R* (red symbols).

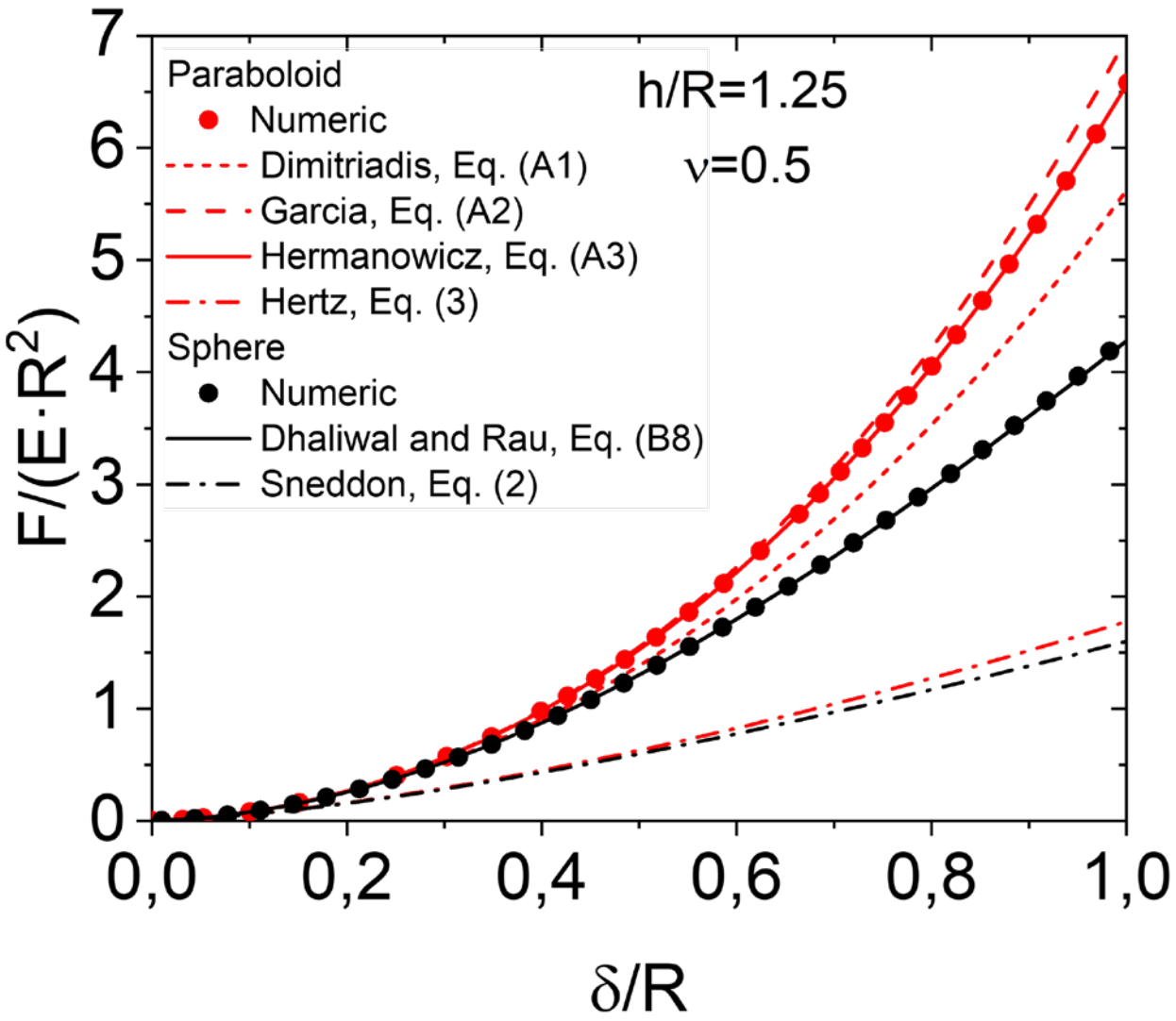

**Figure 2.** (Black and red symbols) Numerically calculated force vs indentation curves corresponding, respectively, to a spherical indenter of radius R and to a paraboloid indenter of radius of curvature R, obtained by solving numerically the corresponding contact mechanics integral equation (see Materials and Methods). (Red and black dot-dashed lines) Predictions of Sneddon's and Hertz's models for a spherical and a paraboloid tip on a semi-infinite film (Eqs. (2) and (3), respectively). (Red short-dashed, long-dashed and continuous lines) Predictions of paraboloid tip models with bottom effect correction (Eqs. (A1)-(A3), corresponding to Dimitriadis et al. [5], Garcia et al. [16], and Hermanowicz [20], respectively). (Black continuous line) Prediction of Dhaliwal and Rau's truncated analytical solution for a spherical indenter with BEC [22] (Eq. (B8)).

Numerical results show that the paraboloid tip is a good approximation of a spherical tip for only relatively small indentations, specifically up to 30-40% of the tip radius in the present case, as previously mentioned.

Sneddon's model for a spherical indenter [13], [14], [15] (black dot-dashed line in Figure 2)

$$\begin{cases} F_{Sned} = \dfrac{E}{1-\nu^2}\left[\dfrac{1}{2}\left(a^2+R^2\right)\cdot\ln\left(\dfrac{R+a}{R-a}\right) - aR\right] \\ \delta = \dfrac{a}{2}\cdot\ln\left(\dfrac{R+a}{R-a}\right) \end{cases} \quad 0<a<R \tag{1}$$

or its equivalent explicit phenomenological expression [28]

$$F_{Sned} = \frac{E}{1-\nu^2}\frac{4\sqrt{R}}{3}\delta^{3/2}\left(1-\frac{1}{10}\frac{\delta}{R}-\frac{1}{840}\left(\frac{\delta}{R}\right)^2+\frac{11}{15120}\left(\frac{\delta}{R}\right)^3+\frac{1357}{6652800}\left(\frac{\delta}{R}\right)^4\right) \tag{2}$$

and Hertz's formula for a paraboloid indenter (red dot-dashed line in Figure 2)

$$F_{Hertz} = \frac{4}{3}\frac{E}{\left(1-\nu^2\right)}\sqrt{R}\delta^{3/2} \tag{3}$$

largely deviate from the numerically calculated data due to the relevance of the bottom stiffness effect. Hence, they can only be used to model spherical tip probes on thin films for very small indentations (here $\delta/R$<10%). The analytical paraboloid tip models with BEC of Dimitriadis et al., Eq. (A1) [5], Garcia et al. Eq. (A2) [16] and Hermanowicz, Eq. (A3) [20] (short-dashed, long-dashed and continuous red lines in Figure 2, respectively), describe reasonably well numerically calculated values for the paraboloid tip, with Hermanowicz's model [15], Eq. (A3), being the most accurate model, as it was derived semi-phenomenologically by combining an analytical series expansion analysis with least squares fittings to numerically calculated data for a paraboloid tip. The previously discussed paraboloid models do not accurately describe the spherical tip for moderately large indentations, specifically for indentations larger than 30-40% of the probe radius, as shown in Figure 2. Conversely, the series expansion parametric analytical solution of Dhaliwal and Rau for a spherical indenter [22] (Eq. (B8) in Appendix B, black continuous line in Figure 2) can correctly describe the bottom stiffness effect for a spherical indenter at large indentations with agreement over the entire indentation range. To achieve such a good agreement across the full indentation ranges, the infinite series expansion was truncated to terms up to $O(a/h)^5$ in the force expression and to $O(a/h)^4$ in the indentation expression (see Appendix B), wider than proposed in Ref. [24], where smaller indentations were considered. However, Dhaliwal and Rau's analytical solution does not hold for very thin films at large indentations where $a/h>1$ [22]. Figure 3 shows the results explicitly, comparing the truncated Dhaliwal and Rau analytical solution, Eq. (B8) (black dashed lines), with numerical data calculated for a spherical tip of radius *R over* a wide range of thicknesses, parameterised by the ratio *h/R*, for a Poisson's ratio $\nu$=0.5. The truncated Dhaliwal and Rau's analytical solution, Eq. (B8) (black dashed lines in Figure 3), correctly describes force-indentation curves for typically $h/R>1$, while for $h/R<1$ it covers only indentations for which $a/h<1$, in agreement with Ref. [22]. The truncated solution is within the range of validity, with a relative error below 3.5%, which is acceptable for many purposes in AFM nanomechanical measurements.

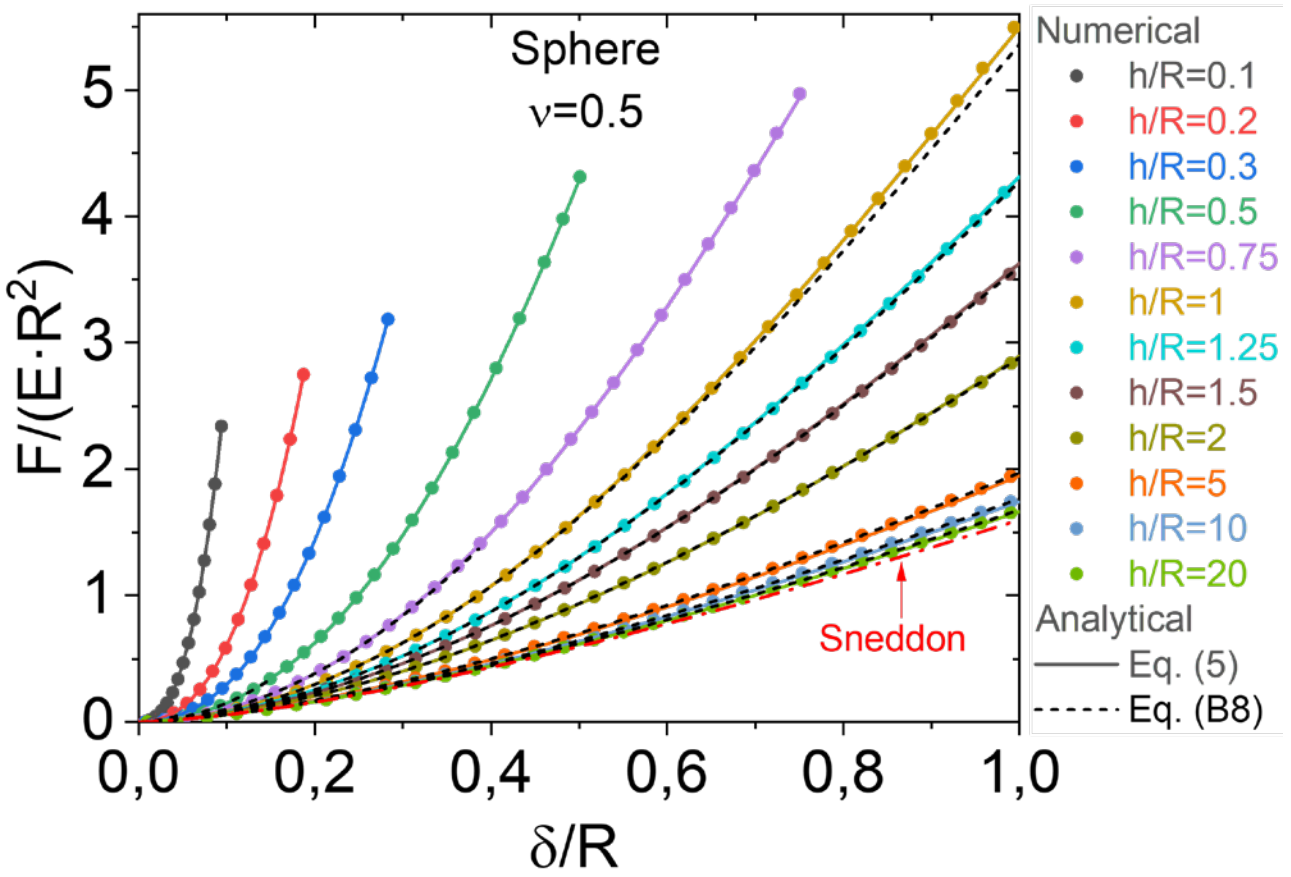

**Figure 3.** (symbols) Dimensionless force-indentation curves calculated numerically by solving the contact mechanics integral equation for different ratios of sample thickness to probe radius, *h/R,* for $\nu$=0.5. (Black dashed lines) Force-indentation curves predicted by the truncated analytical solution of Dhaliwal and Rau [22], Eq(B8). (Colour continuous lines) Force-indentation curves predicted by the analytical model proposed in the present work, Eq. (5).

A simpler parametric expression for Dhaliwal and Rau's truncated analytical solution can also be derived (see Eq. (B9)), which is more tractable and provides reasonable accuracy in these ranges (relative errors below 6%, see Appendix B). This simpler expression can be derived in an independent way by using the methods used by Dimitriadis et al. [5] and Garcia et al. [16], showing that both approaches are equivalent. Still, this simpler expression is only valid for $a/h$<1. To derive an analytical expression valid over the indentation range, i.e. for both $a/h$>1 and $a/h$<1, a phenomenological strategy like the one used in Ref. [20] has been followed. By first deriving the expected functional dependence of the force curve on the ratios $R/h$ and $\delta/R$ by performing a Taylor's series expansion and subsequent inversion of the truncated analytical solution of Dhaliwal and Rau valid for $a/h$<1, Eq. (B8), [22]. Retaining terms up to $O(\delta/R)^{5/2}$ (and the corresponding terms for $R/h$ given by the analytical solution), the following equation is obtained (see Appendix C):

$$F(\delta,R,h,E,\nu)=F_{Sned}(\delta,R,E,\nu)\left\{\begin{array}{l}1+a_1\left(\frac{R}{h}\right)\left(\frac{\delta}{R}\right)^{1/2}+a_2\left(\frac{R}{h}\right)^2\left(\frac{\delta}{R}\right)+\left[a_3\left(\frac{R}{h}\right)+a_4\left(\frac{R}{h}\right)^3\right]\left(\frac{\delta}{R}\right)^{3/2}\\+\left[a_5\left(\frac{R}{h}\right)^2+a_6\left(\frac{R}{h}\right)^4\right]\left(\frac{\delta}{R}\right)^2+\left[a_7\left(\frac{R}{h}\right)+a_8\left(\frac{R}{h}\right)^3+a_9\left(\frac{R}{h}\right)^5\right]\left(\frac{\delta}{R}\right)^{5/2}\end{array}\right\} \quad (4)$$

where the coefficients $a_n$ are functions of only Poisson's ratio, $a_n=a_n(\nu)$. It is assumed that the functional dependence in Eq. (4) is valid for all thicknesses and indentations, with coefficients $a_n$, determined by least squares fitting to the numerically calculated data. For $\nu$=0.5 and the numerical data in Figure 3, which include indentations for both $a/h$<1 and $a/h$>1, the following is obtained (see Appendix C):

$$F(\delta,R,E,h,\nu=0.5)=F_{Sned}(\delta,R,E,\nu=0.5)\times$$
$$\left[\begin{array}{l}1+1.01345\left(\frac{R}{h}\right)\left(\frac{\delta}{R}\right)^{1/2}+2.24003\left(\frac{R}{h}\right)^2\left(\frac{\delta}{R}\right)+\left(-0.470293\left(\frac{R}{h}\right)+0.934118\left(\frac{R}{h}\right)^3\right)\left(\frac{\delta}{R}\right)^{3/2}\\+\left(-0.455595\left(\frac{R}{h}\right)^2-0.124456\left(\frac{R}{h}\right)^4\right)\left(\frac{\delta}{R}\right)^2+\left(0.204097\left(\frac{R}{h}\right)-0.938828\left(\frac{R}{h}\right)^3+0.0291373\left(\frac{R}{h}\right)^5\right)\left(\frac{\delta}{R}\right)^{5/2}\end{array}\right] \quad (5)$$

Figure 3 (continuous colour lines) shows the predictions of Eq. (5) and compares them with the numerically calculated data (symbols), demonstrating that Eq. (5) reproduces the numerical data with a relative error of less than 1.3% for the required ranges of thicknesses and indentations. Alternative coefficients $a_n$ for other values of the Poisson's ratio, $\nu$, are given in Table C1 in Appendix C. Therefore, Eq. (5) for $\nu$=0.5 (or Eq. (4) with the coefficients $a_n$ given in Appendix C for other values of ν) constitutes an accurate representation of the exact numerically calculated data for the case of a spherical tip indenting a linear elastic thin film bottom bonded to a rigid substrate with non-adhesive frictionless contact. It should be noted that, in the limit of thick samples (roughly $h/R>50$) the analytical model, Eq. (5), tends to Sneddon's model, Eq. (2). In contrast, the paraboloid models with BEC, such as those in Eqs. (A1)-(A3), tend to converge to Hertz's model, Eq. (3), in the thick film limit. Although the analytical model has been derived for numerical data satisfying $\delta/R<1$, it holds for larger indentations, at least up to $\delta<\min(h,1.75R)$.

## 3. Comparison of the sphere model with BEC against Sneddon's, Hertz's and the paraboloid tip models at large indentations.

As mentioned before, Sneddon's, Hertz's and paraboloid tip models with BEC cannot describe a sphere indenting a thin elastic film at large indentations (see Figure 2). However, there is a tendency to use these models to interpret force indentation measurements performed with spherical probes, even beyond their range of validity. This tendency arises because these models provide good mathematical fits to the experimental data across the range of indentations under certain conditions. Figures 4a-4c illustrate this point and show fits of Sneddon's Eq. (2), Hertz's Eq. (3), and Hermanowicz's paraboloid with BEC, Eq. (A2), models to the numerical data calculated for a spherical tip (same numerical data as in Figure 3). Continuous black lines correspond to fits performed with the contact point left as a free fitting parameter (replacing $\delta$ with $\delta-z_c$ in the models), while the black dashed lines correspond to fits with the contact point fixed to zero, $z_c=0$, corresponds to the numerical calculations. The fittings to the numerical data are of high quality when the contact point is left as a free parameter for all thicknesses and indentations ($0 \leq \delta \leq min(h,R)$), leading to $R^2$ values exceeding 0.99 (see Figure 4d), even though the models are not supposedto hold in the large indentation regime and thin films. When the contact point is fixed, however, the quality of the fits is much poorer (dashed lines in Figures 4a-4c), reflecting clearly that these models describe different physical conditions and geometries.

Even if the quality of the fits is good, the Young's modulus and contact point obtained with these models differ substantially from the actual Young's modulus and contact point used in the numerical calculations (see Figures 4e and 4f, respectively). For the case of Sneddon's or Hertz's models, the effective Young's modulus shows deviations approaching a factor of 100 in the thin film range $h/R<1$, which is precisely the manifestation of the bottom contact effect. For the case of the paraboloid tip model with BEC (Hermanowicz's model, Eq. (B3) [20]), the effective Young's modulus can be underestimated up to a factor of 0.4 for $h/R\sim 1$.

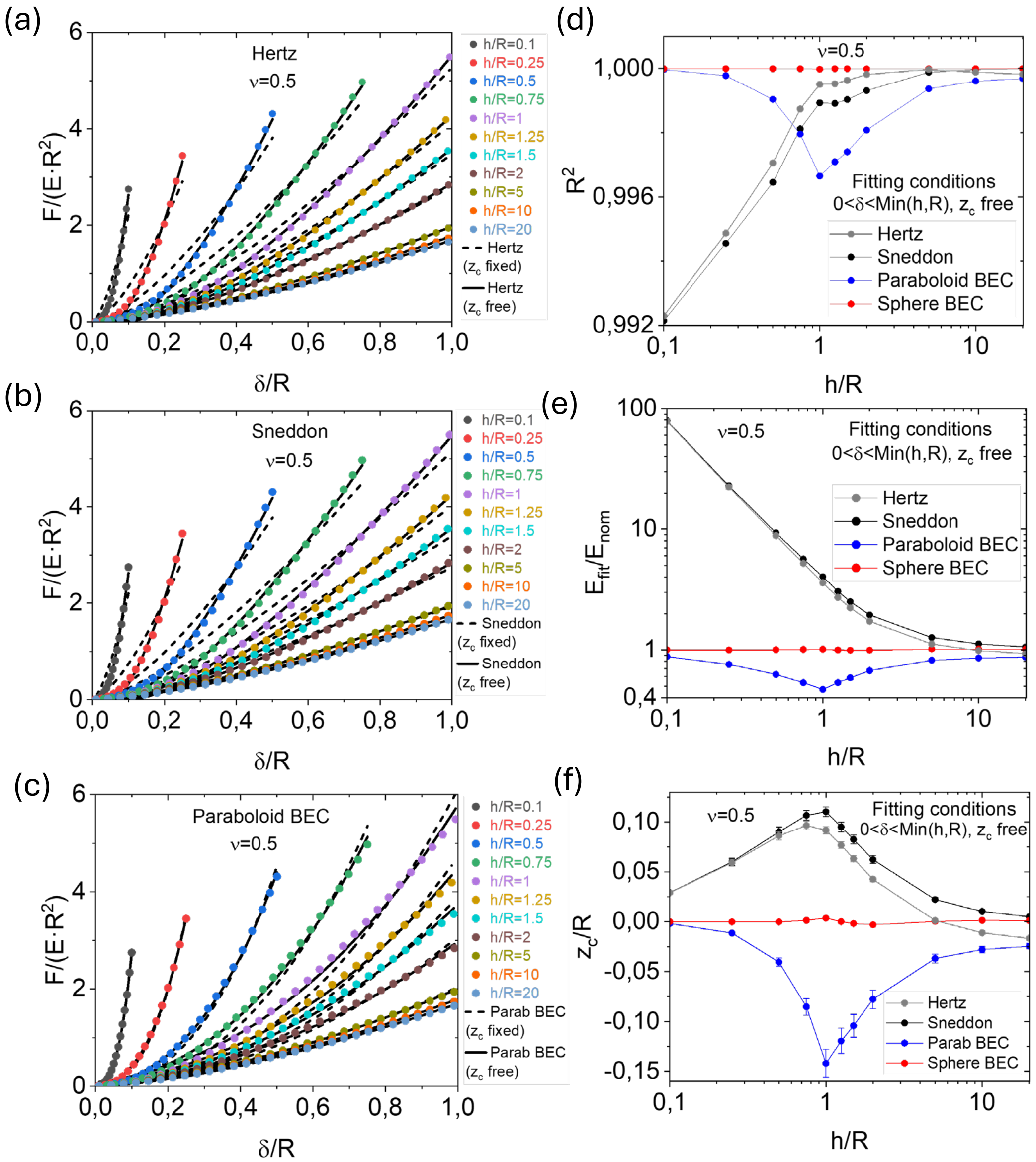

**Figure 4.** (Black dashed and continuous lines) Least-squares fitting of (a) Hertz's, (b) Sneddon's and (c) Hermanowicz's paraboloids with BEC tip models to numerically calculated data for a spherical indenter on a thin film (symbols, same data as in Figure 3) by leaving the contact point fixed to 0 or as a free fitting parameter, respectively. The fits are performed across the whole range of indentations, $0<\delta<Min(h,R)$. (c) $R^2$ values for the fits performed with the different models shown in (a)-(c) and for the sphere with the BEC model presented here. (e) Ratio of fitted Young's modulus to the nominal value used in the numerical calculations for the models considered in (a)-(c) and for the sphere model with BEC presented here (grey, black, blue and red symbols, respectively). (f) Idem to (e) but for the contact point normalised to the radius.

Moreover, when Sneddon's or Hertz's models are used, the contact point is overestimated by up to 10% of the tip radius, while for paraboloid with BEC models, it is underestimated by up to 15% of the probe radius. Note that the contact point is a parameter with physical information, as it is used to determine the true topography of the sample by accounting for the indentation at the force setpoint [29]; the topography would therefore be overestimated or underestimated by the same amount. Therefore, even if existing analytical models can provide fits to the numerical data corresponding to a sphere indenter with values of $R^2$ > 0.99, this fact alone does not guarantee a meaningful physical description of the intrinsic micro/nanomechanical (and topographic) properties of the material. The sphere model with BEC presented here (Eq. (4)), however, provides an excellent fit to the numerical data in the full range of indentations and sample thicknesses with the highest $R^2$ (see Figure 4d, red symbols). Young's moduli are almost identical to the nominal value used in the numerical calculations (relative error less than 1.3%, Figure 4e, red symbols) with null contact points (Figure 4f, red symbols), considered independently of the film thickness and range of indentations, as intended. Physical meaningful parameters can be obtained with Sneddon's, Hertz's or the paraboloid model if they are used within their ranges of validity for a spherical probe, typically for relatively small indentations (see Figure 2). However, restricting the fits to short ranges can significantly reduce their accuracy; furthermore, it can introduce some arbitrariness in the selection of the fitting range. In experiments, a short range can also induce relatively large uncertainties in the determination of the contact point due to other short-range forces that can appear at small distances from the sample. These aspects become irrelevant when the analytical model proposed here, Eq. (4), is used at large indentations, thus offering a robust and reliable approach to determine and map the mechanical properties of elastic thin films at the micro and nanoscale with spherical tip AFM probes.

## 4. Experimental validation of the sphere model with BEC at large indentations on thin PAA hydrogel films.

To validate experimentally the analytical model presented here, Eq. (5), AFM force-indentation measurements at large indentations were performed on thin polyacrylamide (PAA) hydrogel films with an electron beam deposited spherical tip probe with radius $R$=2 μm (Figure 1b). PAA hydrogels are widely used in mechanobiology to study the sensitivity of cells to the substrate stiffness [30], [31], [32] and [33]. The Materials and Methods section details how the PAA films were prepared [34], [35] and how the force indentation measurements were performed and quantified to obtain the local Young's modulus. The PAA composition was chosen to achieve a bulk nominal Young's modulus of ~3 kPa [36], [37], [38] and [39]. The first sample analysed consisted of a PAA thin film of variable thickness, with thickness ranging from ~4 µm to ~200 µm ($h/R$~2-100) (see Figures 5a and 5b and Materials and Methods). Low resolution force indentation maps (20x20) have been measured on different 100x100 µm$^2$ areas (see Materials and Methods). Figure 5c shows one example of a whole set of 20x20 raw deflection versus piezo extension curves acquired at a distance around 1 mm from the film edge where the thickness varies from $h$~9 µm to 11 µm, i.e. $h/R$~4.5-5.5 (see true topography image and height histogram in Figure 5d). Figure 5e shows the calibrated force-indentation curves (black lines), together with the fitted curves corresponding to the sphere model with BEC, Eq. (5) (red lines) (see Materials and Methods). Fits have been performed in the whole indentation range, which in the present case reached indentations up to $\delta_{max}$~3.5 µm (i.e. $\delta/R$~1.75). Figures 5f, 5g and 5h show maps of the Young's modulus, contact point variation and $R^2$ values of the fits. Figures 5i, 5j and 5k show the corresponding histogram representation (red bars). The mean Young's modulus obtained is $E_{sphere}$=4.8±0.3 kPa. The sphere model with BEC provides excellent fits to the experimental data in the whole range of indentation, giving a $R^2$~0.99985±0.00008. Figure 5l illustrates the excellent quality of the fits by plotting an individual experimental curve (black line) corresponding to a single pixel together with the fitted curve (red line) (see all individual fitted curves from this experiment in the Supporting Information). The sphere model is practically indistinguishable from the experimental curve in the whole range of indentations, showing a relatively uniform distribution of errors (Figure 5m) with an absolute error with a standard deviation of just $\sigma$=0.2 nN. The excellent agreement of the model with the experimental data provides its experimental validation. For comparison, we have also analysed the same experimental force curve map using Sneddon's sphere model, Eq.

(2), and Hermanowicz's paraboloid model with BEC correction, Eq. (A3), over the entire indentation range. The values extracted from the fits are shown by the grey and blue bars in the histogram plots in Figures 5i, 5j and 5k, and the grey and blue lines in Figures 5l and 5m, respectively. As anticipated theoretically in section 3, these models also provide good fits to the experimental data when the contact point is left as a free parameter, although they are slightly worse than the sphere with BEC model: $R^2_{Sneddon}$~0.9991±0.002, $R^2_{parab}$~0.9992±0.0003, $\sigma_{Sneddon}$=0.6 nN and $\sigma_{parab}$=0.6 nN.

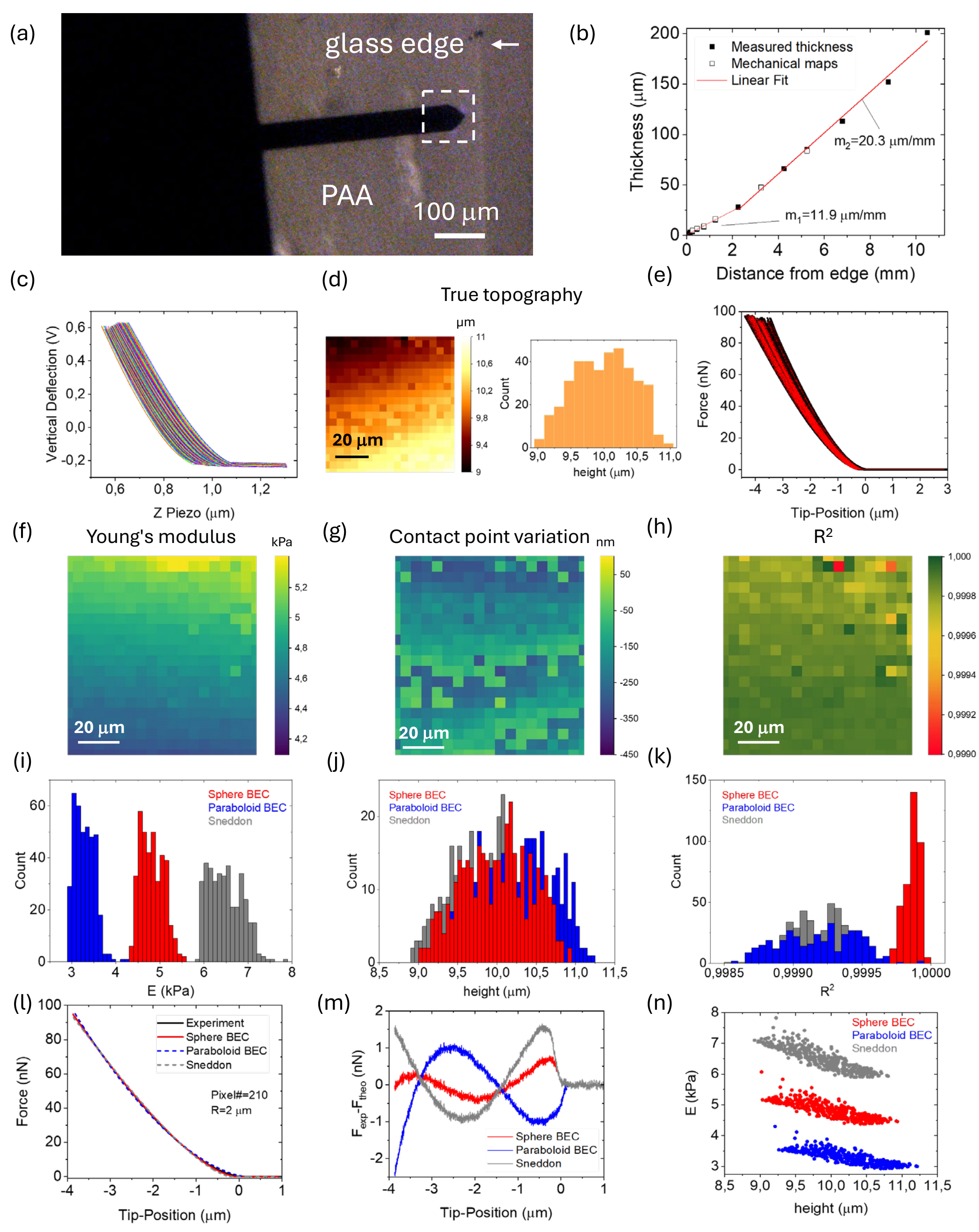

**Figure 5.** (a) Optical microscopy image of the AFM probe located close to the edge of the PAA thin film. The highlighted dashed square represents an area of 100x100 μm$^2$ that was probed by the AFM. (b) (solid symbols) Thickness of the PAA film as a function of distance to the film edge. At the macroscale, the thickness varies at a rate of ~11.9 μm/mm initially and at a slightly faster rate of ~20.3 μm/mm later. (empty symbols) Thicknesses and approximate positions of the regions analysed with the AFM. (c) Raw deflection-piezo extension curves (20x20) measured on a 100x100 μm$^2$ area at ~1 mm awayfrom the film edge. (d) True topography obtained after the data analysis, together with the height distribution histogram. (e) Calibrated force-indentation curves (black lines), together with the curves fitted with the sphere BEC model proposed in the present work (red lines). (f), (g) and (h) Maps of the Young's modulus, contact point variation and $R^2$ of the fits obtained from the fitted curves in (e), respectively. (i), (j) and (k) Histograms of the extracted Young's modulus, true thicknesses and $R^2$ of the fits obtained from (f), (g) and (h), respectively. (l) (Black and red lines) Experimental sphere with BEC fitted force-indentation curves corresponding to a single pixel (#210). For comparison, fitted curves were obtained by using Sneddon's and the paraboloid with BEC models (grey and blue dashed lines). (m) Absolute error of the fitted curves in (l) for the different models as a function of indentation. (n) Extracted values of the Young's modulus as a function of the extracted true topography for the different models considered. Experimental parameters: biosphere B2000-CONT probe with $k$=0.49 N/m, $R$=2 μm, approach velocity $v$=100 μm/s, $Z_{length}$=8 μm, pixel time 180 ms and points per curve 8000.

However, the values of the Young's modulus extracted with these models differ significantly from the ones obtained with the sphere with BEC model; specifically, Sneddon's model (grey bars) gives higher values, $E_{Sneddon}$=6.5±0.4 kPa, while the paraboloid model (blue bars) gives smaller ones, $E_{parab}$=3.3±0.2 kPa, as anticipated theoretically. The extracted contact points also differ among the different models, leading to different estimations of the true thickness of the film, namely, $h_{parab}$=10.2±0.5 μm and $h_{Sneddon}$=9.9±0.4 μm compared to $h_{sphere}$=10.0±0.4 μm (see Figure 5j), again in agreement with the analysis performed in section 3. As stated before, for the present experiment the physically meaningful results are expected to be those corresponding to the sphere model with BEC presented here since a spherical tip probe was used.

**5. Thickness dependence of the Young's modulus of thin PAA hydrogel films.**

The Young's modulus value obtained before $E_{sphere}$=4.8±0.3 kPa differs significantly from the expected nominal value of ~3 kPa for this composition of the PAA hydrogel [31]. In addition, the values seem to vary with the thickness of the film, as observed by comparing the topography and Young's modulus maps in Figures 5d and 5f. This dependency is explicitly illustrated in Figure 5n, where we plot the Young's modulus as a function of the sample thickness. This result suggests that the material's mechanical properties depend intrinsically on the thickness of the material in this thickness range, since its variation with thickness cannot be due to the bottom stiffness effect, which is accounted for by the sphere model. To further investigate the eventual dependence of the Young's modulus on thickness, we have considered five additional regions in the PAA film shown in Figure 5a, moving the tip at different distances from the edge of the bare glass substrate by using the motor sample stage, covering thicknesses from 5 µm to 86 µm. Figure 6a shows the cumulative representation of the Young's modulus measured as a function of the film thickness obtained in the six regions analysed (see data analysis in the Supporting Information). For thicknesses larger than ~15 µm, the Young's modulus approaches a bulk value of ~3.36±0.04 kPa, which is in good agreement with the expected nominal value for this composition [31]. Using Sneddon's model, the value would be ~3.43±0.04 kPa, while with paraboloid BEC models, the value would be ~2.82±0.05 kPa. For thicknesses below ~15 µm, the Young's modulus shows a strong dependence on film thickness, reaching values of up to ~80 kPa for thicknesses around ~6 µm [31]. High quality fits are maintained across the entire range of explored thicknesses (see Figure 6b, red line), where $R^2$ values are almost always larger than 0.999. As before, good fits can also be obtained with Sneddon's and the paraboloid model with BEC (grey and blue symbols in Figure 6b), but the sphere model systematically provides better quality fits from a statistical point of view. From Figure 6a, the trends of the variation in Young's modulus as a function of the thickness are relatively similar for the three models considered, which suggests that it is due to a properly corrected bottom stiffness effect. We have confirmed the dependency of the Young's modulus of PAA films on thickness with independent measurements performed on different samples with uniform thickness and with direct access to the substrate in the imaging region to accurately obtain the sample thickness (see Materials and Methods). The films have a slightly different chemical composition, but, having a similar ratio of monomer (acrylamide) and crosslinker (bisacrylamide) [38], they are expected to have the same nominal bulk Young's modulus, i.e., ~3 kPa. Figure 6a (empty symbols) corresponds to the average results obtained on two sets of uniformly thick samples (one set prepared on glass slides and one set on glass bottom petri dishes to show reproducibility, see Supporting Information). These additional results fully confirm that below a critical thickness (here ~15 μm), the Young's modulus of PAA hydrogel thin films increases sharply with decreasing sample thickness.

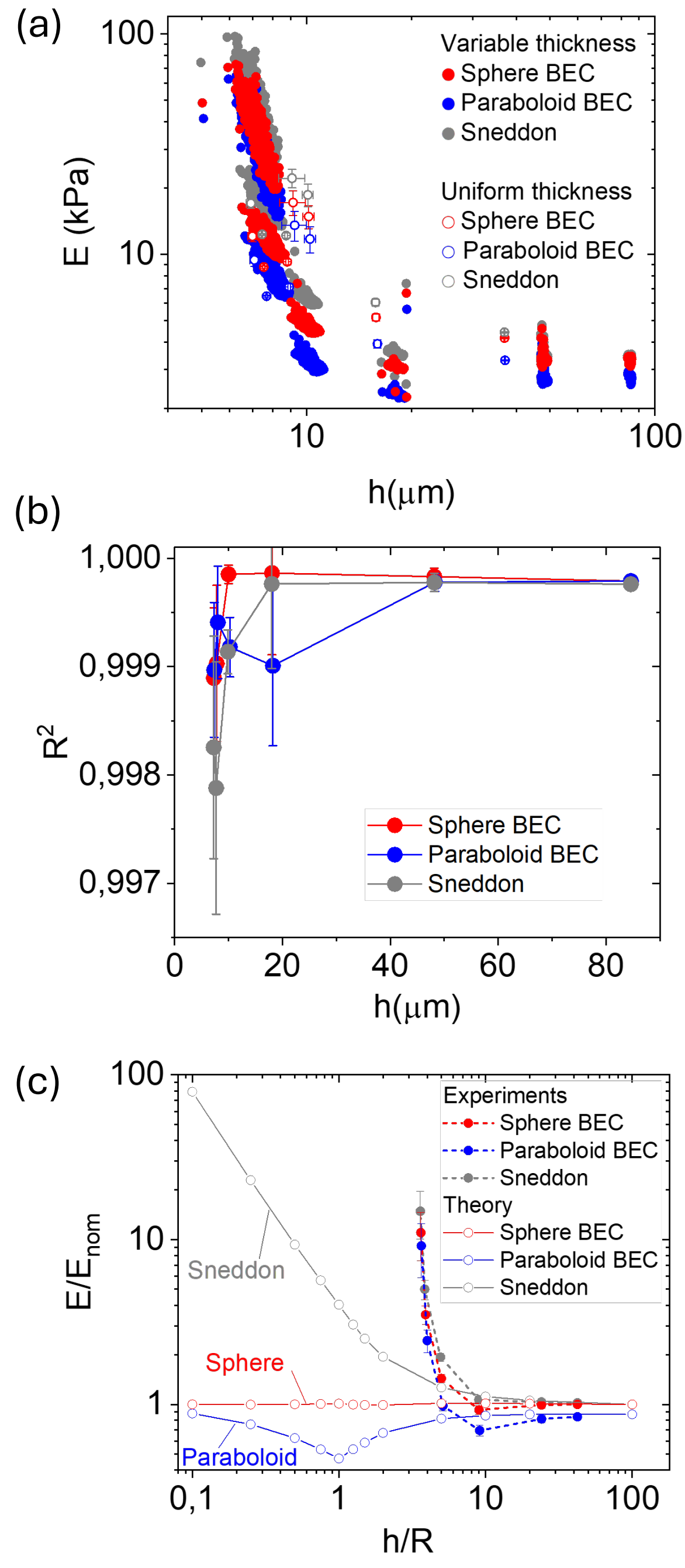

**Figure 6.** (a) (Solid symbols) Cumulative representation of the extracted Young's modulus in the six regions probed on the variable thickness PAA film versus the true sample thickness.

(Empty symbols) Idem, but for uniformly thick additional PAA films. The empty symbols represent the average of the extracted values. (b) Average $R^2$ values of the fits of the different models to the experimental data for the gels with variable thickness. (c) (Solid symbols) Averaged values of the Young's modulus for each region probed in (a) are normalised by the bulk. Young's modulus versus the average thickness normalised by tip radius and represented on a log-log scale. (Empty symbols) Theoretical prediction of the expected behaviour of Young's modulus due to the bottom stiffness effect for the case of Sneddon's sphere and Hermanowicz's paraboloid BEC models (same data as in Figure 4e).

To show that the increase in Young's modulus is not due to the bottom stiffness effect, in Figure 6c we compare the observed experimental behaviour (solid symbols) with the predicted behaviour of the bottom stiffness effect associated with the use of Sneddon's and Hermanowicz's paraboloid BEC models (empty grey and blue symbols, respectively). The solid symbols in Figure 6c represent the average Young's modulus in each region of the variable thickness sample normalised with respect to the bulk value plotted against the thickness normalised to the tip radius. The theoretical data (empty symbols) are the same as those in Figure 4e. The bottom contact effect associated with the use of either Sneddon's model or the paraboloid with BEC model does not follow the trend displayed by the experimental data for $h/R<10$, meaning that the observed variation cannot be attributed to a non-properly corrected bottom stiffness effect, and hence, they reflect an intrinsic variation of the Young's modulus. The increase of the Young's modulus with thickness follows approximately a phenomenological exponential dependency

$$E_{PAA}\left(h\right) = E_{PAA,0} + \alpha e^{-h/h_0} \tag{6}$$

with $E_{PAA,0}$=3.4±0.1 kPa being the bulk value and α=2.3±0.2 GPa and $h_0$=1.5 μm, two phenomenological fitting parameters.

**6. Micromechanical properties of macrophage living cells probed with colloidal AFM probes at large indentations.**

As an application of the sphere model presented here to the case of living cells, we have measured the micromechanical properties of THP1-M0 macrophage cells at large indentations by using a colloidal AFM probe (Figure 1a) with radius $R$=5 μm. Macrophage cells display a very low Young's modulus, in the range of 100s Pa, whose accurate measurement poses important challenges and has attracted recent attention [42], [43]. Figure 7 shows the results of the analysis. Figure 7a shows the topography at a force setpoint $F_{set}$=7 nN, extracted from a 32x32 force map acquired on a 35x25 μm$^2$ area. Figure 7b (black continuous line) shows the cross-section profile along the white dashed line in Figure 7a. For comparison, we also show (red dashed line) the true topography profile of the region of interest (ROI) (black dashed circle in Figure 7a).

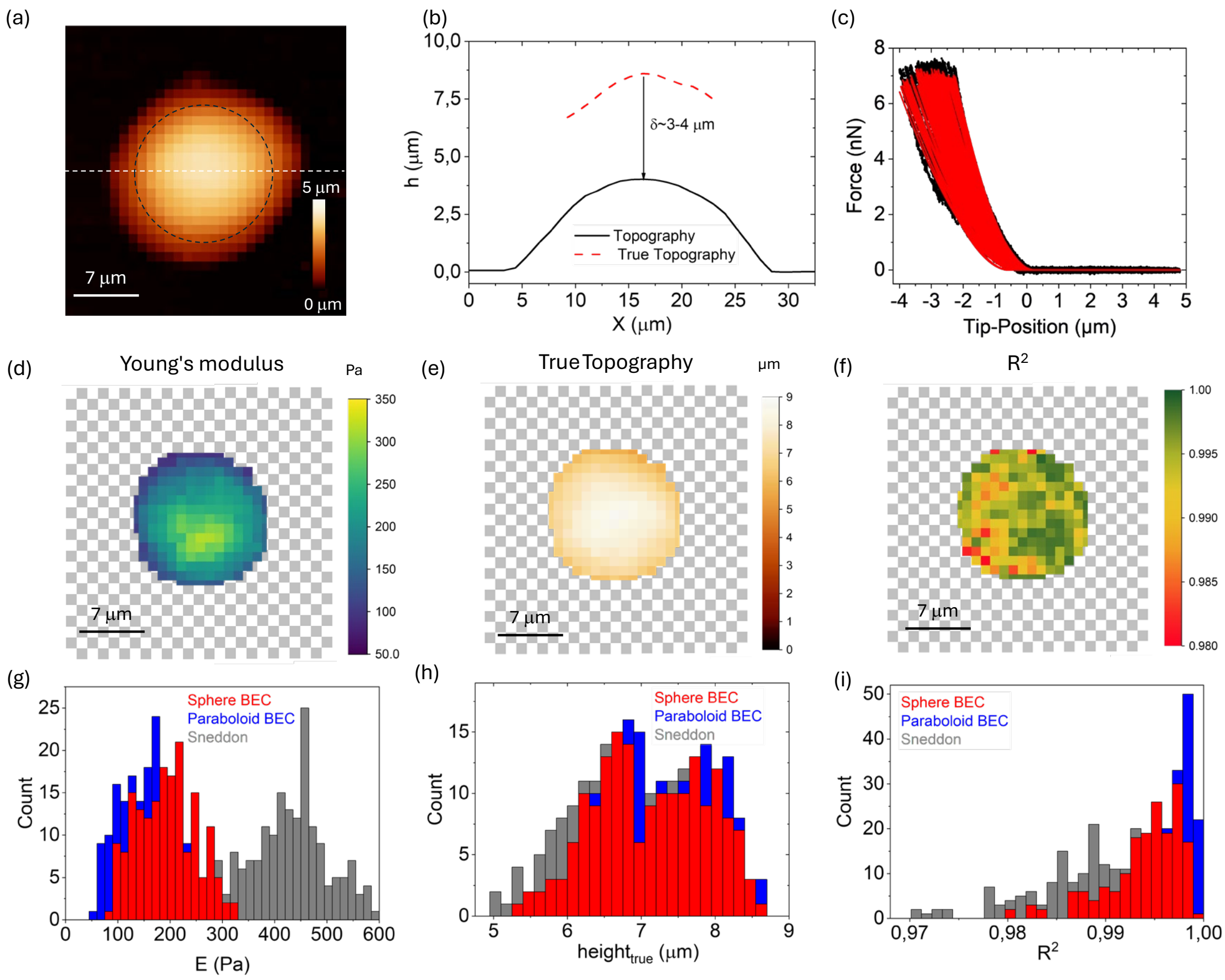

**Figure 7.** (a) The AFM topographic image shows a THP1-M0 macrophage living cell with a force setpoint of $F_{set}$=7 nN, acquired with a colloidal AFM probe of radius $R$=5 μm. (b) (Black line) Cross-section topographic profile along the white dashed line in (a). (Red dashed line) True topography of the ROI (black dashed circle in (a)) obtained after data processing. (c) Fitted curves (red lines) with the sphere model presented in this work to the experimental force indentation curves (black lines). (d), (e) and (f) Maps of the Young's modulus, true topography and $R^2$ of the fits on the ROI. (g), (h) and (i) Young's modulus, height and $R^2$ histograms corresponding to the data in (d), (e) and (f), respectively, as obtained from the sphere model with BEC (red bars), the paraboloid model with BEC (Hermanowickz's) (blue) and Sneddon's (grey bars) models. Experimental parameters: Polystyrene microsphere probe (Novascan) with $k$=0.12 N/m, $R$=5 µm, approach velocity $v$= 40 µm/s, $Z_{length}$=8 µm, pixel time 420 ms, and number of points per curve 8000.

The true height is thus in the range of ~8 µm ($h/R$~1.6), meaning that the force applied can produce large indentations in the range ~3-4 µm ($\delta/R$~0.4-0.5). Figure 7c shows the experimental force-indentation curves (black lines) with the fits using the sphere model with BEC presented here (red lines), again showing an excellent agreement over the entire range of indentations considered, $R^2_{sphere}$~0.994±0.004. Figures 7d, 7e and 7f show maps of the Young's modulus, true topography and $R^2$ of the fits obtained from the fits with the sphere model with BEC. As anticipated before, the Young's modulus obtained is very small in the 100s Pa, $E_{sphere}$=193±56 Pa. This very soft nature explains why small forces (7 nN) can produce large deformations in these cells, which causes the topography to differ significantly from the true topography (Figure 7b). For comparison, we also show in Figures 7g, 7h and 7i the results obtained by using Sneddon's (grey bars) and the paraboloid with BEC (Hermanowicz's) (blue bars) models. Due to the relevance of the bottom stiffness effects, Sneddon's model predicts a significantly larger Young's modulus, $E_{Sneddon}$=428±73 Pa, more than double the value obtained with the sphere model with BEC. Its stiffer nature means that the true topography predicted is slightly smaller (see Figure 7h). In this case, the quality of the fits was poorer than for the sphere model with BEC, $R^2_{Sneddon}$~0.988±0.006. Concerning the paraboloid model with BEC, it predicts a Young's modulus slightly smaller (25% smaller, $E_{parab}$=156±53 Pa) with a relatively similar quality of the fits $R^2_{parab}$~0.996±0.003. We note that even if the maximum indentations are in the range $d_{max}$~$4\ \mu m$, which corresponds to only $d/R$~0.4, just above where paraboloid and sphere models predict different behaviours (see Figure 2), still there is a significant difference in the extracted Young's modulus, with the sphere model being expected to provide the intrinsic value. In future work, a thorough analysis of the mechanical properties of macrophages will be presented using the sphere model introduced here.

**7. Discussion**

An analytical model for quantifying force-indentation curves acquired with spherical tip AFM probes for thin elastic samples on rigid substrates has been presented in Eq. (4), with the coefficients $a_n(\nu)$ depending on Poisson's ratio (see Table C1). For the common case of $\nu$=0.5 the explicit expression is given in Eq. (5). The model is valid for isotropic, homogeneous, and linear thin films of elastic materials bottom-bonded to a rigid substrate, with frictionless and non-adhesive probe-sample contact. The derivation of the model was based on the series expansion analytical solution of Dhaliwal and Rau [22], which is valid for a/h<1, and was phenomenologically generalised to the range a/h>1 using numerically calculated results obtained from the contact mechanics integral equation, Eq. (8). The model reproduces the numerically calculated data accurately, with a relative error less than 1.3% (see Figure 3). The proposed model correctly describes force indentation curves on thin elastic films at large indentations, where Hertz's, Sneddon's, and the paraboloid with BEC models lose their theoretical validity. Despite this loss of theoretical support, we have shown that those models still can provide excellent fits to the numerically calculated data for a sphere indenter in the case that the contact point is set as a free parameter (Figure 4d). However, the values of the Young's modulus (and contact point) predicted by these models can be far from the values obtained with the correct sphere model (see Figures 4e and 4f).

The spherical analytical model has been validated experimentally with large force indentation curves measured with a spherical tip probe on PAA hydrogel thin films bonded to a glass substrate. The model describes the force-indentation curves very accurately across the entire range of indentations and thicknesses, leading to very large $R^2$ values ($R^2 > 0.999$) in all cases (see Figure 6b). The accuracy reached with the proposed model is slightly superior to the ones achieved with alternative models like Sneddon's model or paraboloid models with BEC when used in the same indentation range (see Figure 6b). These models still provide satisfactory fits to the experimental data due to the use of the contact point as a free parameter, in agreement with theoretical predictions (see Figure 5l). However, the extracted Young's modulus is overestimated by Sneddon's and Hertz's models, while the paraboloid model with bottom contact effect correction underestimates it compared to the values obtained with the sphere model presented here (see Figure 5i). The Young's modulus of the PAA hydrogel thin films is intrinsically dependent on the film thickness for thicknesses below 15 µm. For $h>15$ µm, the Young's modulus is independent of film thickness and is ~3.4 kPa (~2.8 kPa if evaluated with Hertz's model) in agreement with reported data in the literature for bulk PAA hydrogels of similar composition [38]. For $h<15$ µm, the Young's modulus shows an abrupt increase, nearly exponential, Eq. (6). This increase in the Young's modulus does not come from the bottom stiffness effect, which is accounted for by the sphere model presented here. This abrupt increase is not due to non-linear elastic effects. Non-linear elastic effects on PAA hydrogels have been analysed in the literature and are explained in terms of neo-Hookean models [23], [26] and [38]. In general, they lead to relatively small corrections for soft and thin film PAA hydrogels (from 7% to 10%). These non-linear corrections are, in fact, smaller if evaluated with the proper geometric tip model. Indeed, using the sphere model presented here to re-analyse the results reported in Ref. [23], obtained with microsphere gravimetric measurements, rather than using Dimiatridis's model for a paraboloid probe [5], shows that non-linear effects reduce to just a 3% correction at most (see Supporting Information), which is quite consistent with the relative absence of non-linear effects observed in our experiments. Non-linear effects were also reported to be insignificant for these hydrogels in Ref. [25]. Therefore, the Young's modulus variation with thickness reflects an intrinsic variation of the mechanical properties of the PAA hydrogel. A similar abrupt increase in Young's modulus of PAA hydrogels for thicknesses around ~15 µm was reported in Ref. [31], where microsphere gravimetric measurements were performed, although, in this case, the effect was attributed to the bottom contact effect, since the quantitative analysis was carried out by using Hertz's model. In Ref. [25], an abrupt increase in the Young's modulus of PAA hydrogel thin films was also observed, this time for thicknesses below ~50 μm. In the latter case, the effect was attributed to an intrinsic variation of the material properties, as Hertz's model, which corrected for both large indentations and bottom contact effects (using a paraboloid model), was used. The larger critical thickness found in this later study could be due to the type of cross-linking used (UV in Ref. [25] versus chemical in our case) or to differences in composition and sample preparation. It is unclear, why the Young's modulus of PAA hydrogels may depend on the thickness for very small thicknesses. The relationship between the structural, solvent and mechanical properties of hydrogels is still the subject of intense research [34], [35], [39]. Models for bulk materials based on equilibrium swelling, rubberlike elasticity, and solute transport have been developed to quantitatively predict physical properties like the swelling ratio, the elastic modulus and solutes' diffusion coefficients within the hydrogels from a set of polymer-specific identity parameters and synthesis-controlled parameters [39]. For thin film materials, where interfaces may play a significant role, corresponding models have not yet been reported. Therefore, at present, it is difficult to ascertain what interfacial phenomena are responsible for drastically altering the intrinsic mechanical properties of very thin film PAA hydrogels. A possibility could be that the effect is a consequence of a size-dependent porosity and swelling [44]. It has been shown that the swelling of micrometre PAA posts ($h$ > 100 μm) depends on their size and shape [45]. Furthermore, in thick PAA films ($h$ > 1 mm), it has been found that during equilibration in PBS the local Young's modulus decreases as swelling increases [46], confirming the correlation between these two physical parameters. Further research is necessary to clarify this aspect. The Young's modulus of PAA hydrogels depends on thickness, which may require a reinterpretation of the experiments that reported the sensitivity of some cells to the substrate stiffness [30], [32]. In these mechanobiology studies, it was observed that cells were sensitive to the thickness of the hydrogels, which was interpreted as a consequence of the bottom contact effect since cells probed areas were larger than the hydrogels' thickness. In these studies, then, it was assumed implicitly that the PAA hydrogels presented a Young's modulus independent from film thickness. However, direct measurements of the Young's modulus of PAA hydrogel thin films at that time were inconclusive due to the low value of the Young's modulus (in the kPa range), the delicate nature, and the lack of accurate models to model the bottom contact effect for spherical tip probes [23], [31], [36], and [37]. The sphere model developed here shows that if hydrogel films are thinner than a critical thickness, the mechanical properties probed by the cells are affected by both the bottom stiffness effect and the intrinsic variation of the mechanical properties of the thin film.

Measurements performed on living macrophage cells also experimentally validated the model presented (Figure 7). Macrophage cells are extremely soft cells, with Young's modulus in the 100's Pa range, and for that reason small forces may produce large indentations, making it necessary to use models that correctly describe the geometry of the measuring probe. So far, experiments performed on this type of cell [42], [47] have been analysed using a parabolic model with BEC. We have shown that the use of the sphere model with BEC correction predicts slightly larger Young's modulus, around ~25% larger.

As we have mentioned before, the sphere model presented here is valid for isotropic, homogeneous, and linear thin film elastic materials that are bottom-bonded to a rigid substrate, with frictionless, non-adhesive probe-sample contact. Generalising it to cover other situations of interest is beyond the scope of this work. Here, we only mention that the model can be generalised relatively easily for semi-infinite compliant substrates. To this end, one should consider in Eq. (9) the function $C(X)$ corresponding to a compliant semi-infinite substrate (see Ref. [27]) and use it to numerically solve the contact mechanics integral equation and evaluate Dhaliwal and Ray's analytical solution [22]. In that case, the results would depend on the Poisson's ratio, reduced Young's modulus and thickness of the thin film ($\nu_1$, $E_1^*$ and $h_1/R$, respectively) and, in addition, on the Poisson's ratio of the substrate ($\nu_2$) and the ratio of reduced Young's moduli of the thin film and the substrate $E_1^*/E_2^*$, making the analysis far more complex. The generalisation of the model to include non-bonded bottom conditions, viscoelastic or non-linear elastic materials, or adhesive contacts could also be derived, as has been done for other probe geometries [4], [29], but it is expected to be more complex in the case of spherical geometry with large indentations.

## 8. Conclusions

An analytical explicit model to quantify force indentation curves at large indentations acquired on thin elastic materials with spherical tip AFM probes has been presented. The model overcomes the limitations of existing analytical models at large. indentations (e.g., Sneddon's, Hertz's or paraboloid with BEC models) and reproduces almost exactly the results predicted by the contact mechanics integral equation solved numerically (relative errors below 1.3%). The model has been validated experimentally on thin PAA hydrogel films, where fits to the experimental data lead to values of $R^2$ value greater than 0.999 for large indentation curves. By using this model, we showed that PAA hydrogels have a thickness dependent Young's modulus for film thicknesses below a critical thickness (in the present case, below ~15 µm), an effect not related to the bottom stiffness effect, which is accounted for by the model, nor to non-linear elastic properties, which are small in this material and hence reflect an intrinsic variation of the material's mechanical properties. The model has also been validated with large indentation force measurements performed on living macrophage cells that show very low Young's modulus, in the range of 100s Pa, and hence suffer large deformations even when small forces are applied. The model presented here paves the way for more quantitative and reproducible measurements of the elastic properties of thin elastic films and cells at the micro and nanoscale using AFM force spectroscopy with spherical tips. This approach can also be extended to other techniques using spherical indenters.

## Materials and Methods

*Contact mechanics integral equation for a spherical indenter on a thin film and its numerical resolution.* The contact mechanics integral equation to determine the force versus indentation of an axisymmetric indenter indenting a thin film bottom bonded to a rigid substrate (frictionless non-adhesive contact) can be expressed in the form of a Fredholm integral of the second kind, which is readily amenable to numerical resolution. The equation is formulated in terms of two auxiliary functions, *θ(s)* and *g(r),* associated, respectively, to the surface deformation, *u(r),* and pressure distribution, *p(r)* [27]

$$\theta(s)=\frac{d}{ds}\int_0^s dr\frac{ru(r)}{\sqrt{s^2-r^2}};\quad g(s)=\int_s^\infty dr\frac{rp(r)}{\sqrt{r^2-s^2}} \tag{7}$$

The integral equation to determine the function *g(r)* reads [27]

$$\theta(s)=\frac{2}{\pi}\int_0^a g(r)\left[\int_0^\infty C(kh)\cos(kr)\cos(ks)\,dk\right]dr \tag{8}$$

where *a* is the contact radius, *h* the film thickness, and [27]

$$C(X)=\frac{2(1-\nu^2)}{E}\frac{1-\frac{4}{3-4\nu}X\cdot e^{-2X}-e^{-4X}}{1+\left(3-4\nu+\frac{1}{3-4\nu}+\frac{4}{3-4\nu}X^2\right)e^{-2X}+e^{-4X}} \tag{9}$$

Equation (9) is valid for a thin film bonded to a rigid substrate (see the corresponding expression for the non-rigid substrate case in Ref. [27]). The surface deformation *u(r)* in the region of contact is determined by the shape of the indenter, which for the case of a spherical indenter, reads

$$u(r)=\delta-\left(R-\sqrt{R^2-r^2}\right) \tag{10}$$

where *R* is the sphere radius. Substituting Eq. (10) into Eq. (7) gives

$$\theta(s)=\delta-\frac{1}{2}s\ln\left(\frac{R+s}{R-s}\right)=\delta-s\cdot\operatorname{arctanh}\left(\frac{s}{R}\right) \tag{11}$$

The resolution of Eq. (8) with Eq. (11) is subject to the boundary condition [27]

$$g(a)=0 \tag{12}$$

Since out of the contact region, the pressure field nullifies. Solving the integral equation, Eq. (8), gives the function *g(r)*, from where the applied force is calculated as

$$F=4\int_0^a g(r)\,dr \tag{13}$$

The application of the boundary condition, Eq. (12), provides the dependency of the indentation *δ* on the contact radius *a*. The solution of the contact mechanics integral equation leads to a parametric function (*F(a)*, *δ(a)*) in terms of the contact radius, *a*. We have solved Eq. (8)

numerically with Eq. (11), being subject to the boundary condition in Eq. (12) by using Mathematica 14.2 from Wolfram. The Mathematica notebook is available as Supporting Information. In the numerical resolution, we used the dimensionless variables introduced in Ref. [27].

$$\Pi = \frac{2FR\left(1-\nu^2\right)}{a^3 E}; \quad \Delta = \frac{\delta R}{a^2}; \quad \tau = \frac{h}{a} \tag{14}$$

With these variables the solution for a spherical indenter depends only on the ratio *h/R* and Poisson's ratio *v* so that, for each pair of values of *h/R* and *v*, a normalised force indentation curve needs to be calculated.

The calculations for spherical indenters, shown in Figures 2 and 3, have been calculated in this way. A similar procedure can be followed for the case of a paraboloid tip. This case was explicitly treated in Ref. [27]. The only difference with the sphere case is that for a paraboloid tip Eqs. (10) and (11) read

$$u(r) = \delta - \frac{r^2}{2R} \tag{10'}$$

and

$$\theta(s) = \delta - \frac{s^2}{R} \tag{11'}$$

The same dimensionless variables in Eq. (14) are also used. In this case, the model depends only on Poisson's ratio *v*, so that only one normalised force indentation curve needs to be calculated numerically for each value of the Poisson's ratio, which can be used to derive force indentation curves for all values of *h/R*. We have solved Eq. (8) with Eq. (11') subject to the boundary condition in Eq. (12) numerically also by means of Mathematica 14.2 from Wolfram. The Mathematica notebook is available as Supporting Information. The data in Figure 2 for the paraboloid tip were calculated in this way.

*Preparation of PAA hydrogel thin films*. PAA hydrogels with a nominal Young's modulus of ~3 kPa were used in this research. Two slightly different protocols were used, one for the films of variable thickness and one for the films with uniform height. The protocols were developed by two different groups collaborating in this research [45], [48], who validated the Young's modulus of the two formulations on bulk samples by independent AFM and nanoindentation measurements. Prior to hydrogel precursors' deposition, all glass surfaces used in the sample preparation were previously cleaned by a series of 30 min sonications in the following solutions: tap water with soap, tap water, distilled water, 1 mM ethylenediaminetetraacetic acid (EDTA) solution, 70% ethanol and 96% ethanol (all reagents from Sigma-Aldrich). Variable thickness samples. Hydrogels were prepared by copolymerising acrylamide and bis-acrylamide (BioRad) on silanised glass substrates. First, clean glass slides (75mm x 25 mm; Deltalab) were silanised with a 50% (v/v) solution of 3-(aminopropyl)trimethoxysilane (APTMS) (Sigma-Aldrich) in Milli-Q water for 5 min, followed by incubation in Milli-Q water for 30 min on an orbital shaker. Substrates were then rinsed several times with water and incubated with 0.5% (v/v) glutaraldehyde solution (Sigma-Aldrich) in water for an additional 30 min. After rinsing again with water, the substrates were dried with $N_2$ and stored in a vacuum desiccator until use. A solution of 7.5% acrylamide, 0.05% bis-acrylamide, 0.05% ammonium persulphate (APS) (Serva) and 0.15% tetramethylethylenediamine (TEMED) (Sigma-Aldrich) in water was used as the precursors' solution. To produce the variable height, a PDMS spacer (~200 µm) was placed on the silanised glass slide, then the precursors' solution was cast next to it, covered with a clean squared glass coverslip (size: 24x24 mm$^2$), and allowed crosslinking for 2 hours. After crosslinking, hydrogels were demolded by gently removing the glass cover and then stored in PBS at 4ºC until use.

Constant thickness samples. Two sets of samples were prepared on glass slides (Deltalab) and glass-bottom petri dishes (MatTek). Substrates were silanised by incubating them in a solution containing 3-(trimethoxysilyl)propyl methacrylate (Bind-Silane, Sigma Aldrich) for 30 min (7.14%/7.14% - Bind-Silane/acetic acid in ethanol 96%) (Sigma-Aldrich), then rinsed twice with 96% ethanol and left to dry completely. A solution of 6.2% acrylamide, 0.044% bisacrylamide (both from BioRad), 0.05% APS and 0.05% TEMED (both from Sigma-Aldrich) in PBS 1x was used as the precursors' solution. Hydrogels of different thicknesses were prepared by casting varying volumes of the precursors' solution (12 µL, 2.4 µL and 0.6 µL) on the silanised substrates and covering the drops with glass coverslips of different diameters (12 mm and 24 mm for thinner films), thus spreading different volume of the PAA solution on a certain area. After 1h incubation at room temperature, the PAA solution crosslinked into a gel. 1x PBS was then added to samples and left for several minutes (up to 30 min), allowing for easier detachment of the coverslip from the gel. Samples were stored in PBS at 4 ºC until use. Prior to AFM imaging, all samples were allowed to reach room temperature.

*Thickness determination of the hydrogel films.* To determine the thickness of the uniform samples, we scratched the different PAA films with syringe needles to expose the glass substrate. The film thickness was measured with the AFM by imaging an area of the sample that contained the scratch. For samples thicker than the maximum z-piezo extension (15 µm), the thickness was measured using the AFM z-motor. In this case, the tip was first approached to a region of the sample with exposed glass using a small setpoint (5 nN), and then the cantilever was retracted using the z-motor while maintaining a fixed separation distance of 300 µm. The cantilever was then moved to an area covered by the PAA film, re-approached at 5 nN and retracted again from the same separation distance (300 µm) using the z-motor. The difference in the z-motor position readings between the glass and the sample gave the sample height. We found that the uncertainty of the z-motor movement depends on the retraction distance. The 300 µm distance was chosen as the one giving the smallest error (maximum error: ± 500 nm), estimated by repeated retractions after approach at a 5 nN force.

*Cell sample preparation.* THP-1 cells (human acute monocytic leukaemia cell line, ATCC-TIB-202; ATCC) were cultured in RPMI-1640 medium (Gibco, Thermofisher, Ref. 11835063) supplemented with 10% foetal bovine serum (FBS; Life Technologies, Ref. 10270106), 1% penicillin-streptomycin (P/S; Gibco, Thermofisher, Ref. 15140122), and 0.05 mM β-mercaptoethanol (BME; Gibco, Thermofisher, Ref. 31350010). Cells were maintained in suspension in T-75 flasks at 37 °C in a humidified incubator with 5% $CO_2$. Macrophage differentiation was performed in the previously described culture medium. Cells were seeded at a density of $5 \times 10^5$ cells per dish onto FluoroDish glass-bottom culture dishes (35 mm diameter; World Precision Instruments). Differentiation into the unpolarised macrophage phenotype (M0) was induced by incubation with 20 ng/mL phorbol 12-myristate 13-acetate (PMA; Sigma Aldrich, Ref. P1585) for 48 h.

*Force indentation measurements.* Force curve maps have been acquired by means of a commercial AFM system (Nanowizard 4, JPK-Bruker) mounted on an Eclipse Nikon optical microscope by using the QI mode. Force curve maps on hydrogels have been acquired in PBS at the speed of 100 μm/s. We did not find changes in the photodetector voltage vs z piezo evolution during the approach at lower velocities. Datasets contain 25x25 (samples with constant height) and 20x20 force curves (samples with variable height) collected in areas of 100x100 $\mu m^2$. Z-length was always chosen to be between 8 and 9 μm, ensuring that the entire curve's path was clearly recognisable from the horizontal/far-from-contact region to the mechanical deformation. We used biosphere B2000-CONT probes consisting of rectangular cantilevers with diamond-like carbon spherical tip apex of $R$=2 μm electron beam deposited (Nanotools). Figure 1b shows a scanning electron microscopy image of one of the probes used. The spring constant was in the range of 0.3-0.5 N/m, as determined by the thermal noise method. Relatively large force setpoints were set during the experiments to ensure the large indentation regime ($\delta/R$>0.5), was reached. Force curve maps on cells have been acquired using colloidal probes consisting of polystyrene microspheres with a radius of 5 μm (Novascan) (one of these probes is shown in Figure 1a). We used cantilevers with a nominal spring constant of 0.12 N/m, as determined via the thermal noise method prior to the experiment. Force curves have been acquired with a ramp size of 8 μm and an approach velocity of $v$=40 $\mu m\ s^{-1}$. Relatively high force setpoints (7 nN) were used during the experiments to ensure large indentations were reached. Data sets contain 30×30 force curves in an area of 35×35 μm².

*Force curve calibration and quantification*. The calibration and quantification of the force curve maps to extract the Young's modulus maps have been performed with our custom developed software tool Mechanotool, to be presented elsewhere. The data processing is as follows. First, the optical lever sensitivity (OLS) is determined from the contact part of the force curves acquired on the exposed glass substrate (or on an independent bare glass slide if the substrate is not accessible). In the present work the substrate region and the region of interest (ROI) have been segmented manually using a polygon segmentation tool, avoiding regions with low quality curves (e.g., cell or scratch edges). For simplicity, we have not considered the refinements that are sometimes necessary when using colloidal probes, as discussed in Ref. [11]. The baseline of the force curve was determined automatically with no user intervention using a sliding window algorithm that iteratively performs linear fittings on increasingly longer segments of the curve in a forward direction until a peak correlation coefficient is found [49]. Likewise, this method was applied in the inverse direction to detect the linear segment of the indentation portion of the curve during OLS calculation. The linear function corresponding to the non-contact part of the curve provides the offset and tilt of the baseline, used to correct the force curves. The first estimation of the contact point consisted of the breakpoint of the linear baseline segment identified in the previous step. This estimation is independent from the physical model and hence can be used as a reference value to compare the different models. From the calibrated force-piezo curves, the force-indentation curves are obtained by subtracting the deflection from the variation in piezo extension. The calibrated force-piezo extension curves are used to determine the sample topography at the force setpoint ($F_{set}$) and, together with the force indentation curves, to determine the true sample topography, correcting for the indentation induced by the force setpoint. The force indentation curves are then fitted by the sphere model, Eq. (5) (or the alternative models, i.e., Sneddon's model, Eq. (2); Hertz's model, Eq. (3); and Hermanowicz's paraboloid with BEC model, Eq. (A2)), leaving both the Young's modulus and the contact point as free parameters (the contact variation point, $z_c$, is introduced into the models by just exchanging $\delta$ for $\delta$-$z_c$). The film thickness is extracted from the true topography calculated previously. Since the true topography depends on the contact point, a refined estimation of the true topography is performed by using the contact point obtained from the fit of the analytical model, and a second fit of the model is performed to extract refined values of the Young's modulus and contact point. In addition, the Mechanotool software provides further information on the intermediate processing steps, which is included in the Supporting Information.

## Appendix

### Appendix A: Analytical models for a paraboloid tip with bottom contact effect correction

The analytical models for a paraboloid tip with bottom contact effect correction of Dimiatridis et al. [5], Garcia et al. [16] and Hermanowicz [20] for $v$=0.5 read, respectively,

$$F_{Dimitriadis} = \frac{4}{3}\frac{E}{1-0.5^2}\sqrt{R}\delta^{3/2}\left(1+1.133\frac{\sqrt{\delta R}}{h}+1.283\left(\frac{\sqrt{\delta R}}{h}\right)^2+0.769\left(\frac{\sqrt{\delta R}}{h}\right)^3+0.0975\left(\frac{\sqrt{\delta R}}{h}\right)^4\right) \tag{A1}$$

$$F_{Garcia} = \frac{4}{3}\frac{E}{1-0.5^2}\sqrt{R}\delta^{3/2}\left(1+1.133\frac{\sqrt{\delta R}}{h}+1.497\left(\frac{\sqrt{\delta R}}{h}\right)^2+1.469\left(\frac{\sqrt{\delta R}}{h}\right)^3+0.755\left(\frac{\sqrt{\delta R}}{h}\right)^4\right) \tag{A2}$$

$$F_{Hermanowick} = \begin{cases} \dfrac{4}{3}\dfrac{E}{\left(1-0.5^2\right)}\sqrt{R}\delta^{3/2}\left(1+1.105\left(\dfrac{R\delta}{h^2}\right)^{1/2}+1.607\left(\dfrac{R\delta}{h^2}\right)+1.602\left(\dfrac{R\delta}{h^2}\right)^{3/2}\right) & \delta \le 0.4\dfrac{h^2}{R} \\ \dfrac{Eh^3}{\left(1-0.5^2\right)R}\begin{pmatrix} 0.616-3.114\left(\dfrac{R\delta}{h^2}\right)^{1/2}+6.693\left(\dfrac{R\delta}{h^2}\right)-7.170\left(\dfrac{R\delta}{h^2}\right)^{3/2} \\ +8.228\left(\dfrac{R\delta}{h^2}\right)^2+\dfrac{\pi}{2}\left(\dfrac{R\delta}{h^2}\right)^3 \end{pmatrix} & \delta > 0.4\dfrac{h^2}{R} \end{cases} \tag{A3}$$

### Appendix B: Dhaliwal and Rau's truncated analytical solution for a spherical indenter on a thin film for *a/h*<1.

The analytical solution derived by Dhaliwal and Rau in Ref. [22], valid for $a/h<1$ for a spherical indenter with bottom effect correction (frictionless non-adhesive contact), is obtained by expanding the cosines in Eq. (8) with Eq. (11), leading to [22]

$$\begin{cases} F=\dfrac{4aE}{1-\nu^2}\left[\sum_{n=0}^{n_{\inf,F}}\dfrac{\delta p_n-q_n}{2n+1}-\dfrac{R}{4}\left\{2-\left(\dfrac{R}{a}-\dfrac{a}{R}\right)\log\left(\dfrac{R+a}{R-a}\right)\right\}\right] \\ \delta=\dfrac{\sum_{n=0}^{n_{\inf,\delta}}q_n+\dfrac{1}{2}a\log\left(\dfrac{R+a}{R-a}\right)}{\sum_{n=0}^{n_{\inf,\delta}}p_n} \qquad a<R, a<h \end{cases} \tag{B1}$$

where

$$\begin{aligned} p_0&=\sum_{n=0}^{n_{\inf}}\left(\frac{a}{h}\right)^n\alpha_0^{(n)} \\ p_m&=\sum_{n=2m+1}^{n_{\inf}}\left(\frac{a}{h}\right)^n\alpha_m^{(n)};\quad m=1,2,3,... \\ q_m&=\sum_{n=2m+1}^{n_{\inf}}\left(\frac{a}{h}\right)^n\beta_m^{(n)};\quad m=0,1,2,... \end{aligned} \tag{B2}$$

with $\alpha_m^{(n)}$ and $\beta_m^{(n)}$ being defined recursively as

$$\begin{aligned} \alpha_0^{(0)}&=1 \\ \alpha_m^{(n)}&=\frac{2}{\pi}\sum_{i=m}^{[(n-1)/2]}\sum_{j=0}^{[(n-1)/2-i-1]}\frac{(-1)^{m+1}\Pi_{k=0}^{k=m-1}(n-k)}{(2i+2j-2m+1)(m!)}\frac{1}{(2i!)}\left[\int_0^\infty u^{2i}C(2u)\right]\alpha_j^{(n-2i-1)};\quad n=1,2,3,... \\ \beta_0^{(1)}&=-\frac{2}{\pi}\int_0^\infty C(2u)\,du \\ \beta_m^{(n)}&=\frac{2}{\pi}\sum_{i=m}^{[(n-1)/2]}\sum_{j=0}^{[(n-1)/2-i-1]}\frac{(-1)^{m+1}\Pi_{k=0}^{k=m-1}(n-k)}{(2i+2j-2m+1)(m!)}\frac{1}{(2i!)}\left[\int_0^\infty u^{2i}C(2u)\right]\beta_j^{(n-2i-1)};\quad n=2,4,6,... \\ \beta_m^{(n)}&=\frac{2}{\pi}\sum_{i=m}^{(n-3)/2}\sum_{j=0}^{[(n-1)/2-i-1]}\frac{(-1)^{m+1}\Pi_{k=0}^{k=m-1}(n-k)}{(2i+2j-2m+1)(m!)}\frac{1}{(2i!)}\left[\int_0^\infty u^{2i}C(2u)\right]\beta_j^{(n-2i-1)} \\ &\quad+\frac{2}{\pi}(-1)^{(n-1)/2+1}\frac{\Pi_{k=0}^{k=2m-1}(n-1-k)}{(2(n-1)/2!)(2m!)}\left[\int_0^\infty u^{2(n-1)/2}C(2u)\right]\gamma_m^{(n)};\quad n=3,5,7,... \end{aligned} \tag{B3}$$

Where

$$\gamma_m^{(n)}=\frac{a}{2(n-2m+1)}\left\{\left[1-\left(\frac{R}{a}\right)^{n-2m+1}\right]\log\left(\frac{R+a}{R-a}\right)+2\sum_{i=1}^{(n-2m+1)/2}\frac{1}{2i-1}\left(\frac{R}{a}\right)^{n-2m-2i+2}\right\} \tag{B4}$$

The integrals in Eq. (B3) involve the function C(X) defined in Eq. (9) and hence depend on Poisson's ratio $\nu$. They must be evaluated numerically. To examine the solution at different orders of approximation in $a/h$, we have substituted the infinite limits of the sums in Eq. (B1) by the integer parameters $n_{inf,F}$ and $n_{inf,\delta}$, for the force and indentation expression, respectively. In this way, one can evaluate the solution up to a certain order in $a/h$ by setting the corresponding value of $n_{inf}$ to the desired order. Note that strictly speaking, Eq. (B1) for the indentation is expressed as a ratio of series expansions in *a/h*, so here, by order, we refer to the order of the series in the numerator and denominator. The computation of the solution has been performed with Mathematica 14.2 from Wolfram. The notebook is available as Supporting Information. Figure B1a shows the normalised indentation $\delta/R$ as a function of the normalised contact radius *a/R,* as predicted by Eq. (B1) for $h/R$=1 and $\nu$=0.5, which represents the limiting case for the validity of the solution in the form of a series expansion for orders in *a/h* from O(2) up to O(7), in the sense mentioned above. For $a/h$<0.6, the O(2) expression for the indentation already provides an accurate description, but for $a/h$>0.6 the best agreement is obtained by the O(4) expression. Orders higher than O(4) do not provide a better agreement for $a/h$>0.6, since Taylor series expansions tend to show deviations at higher orders when evaluated far from the expansion point.

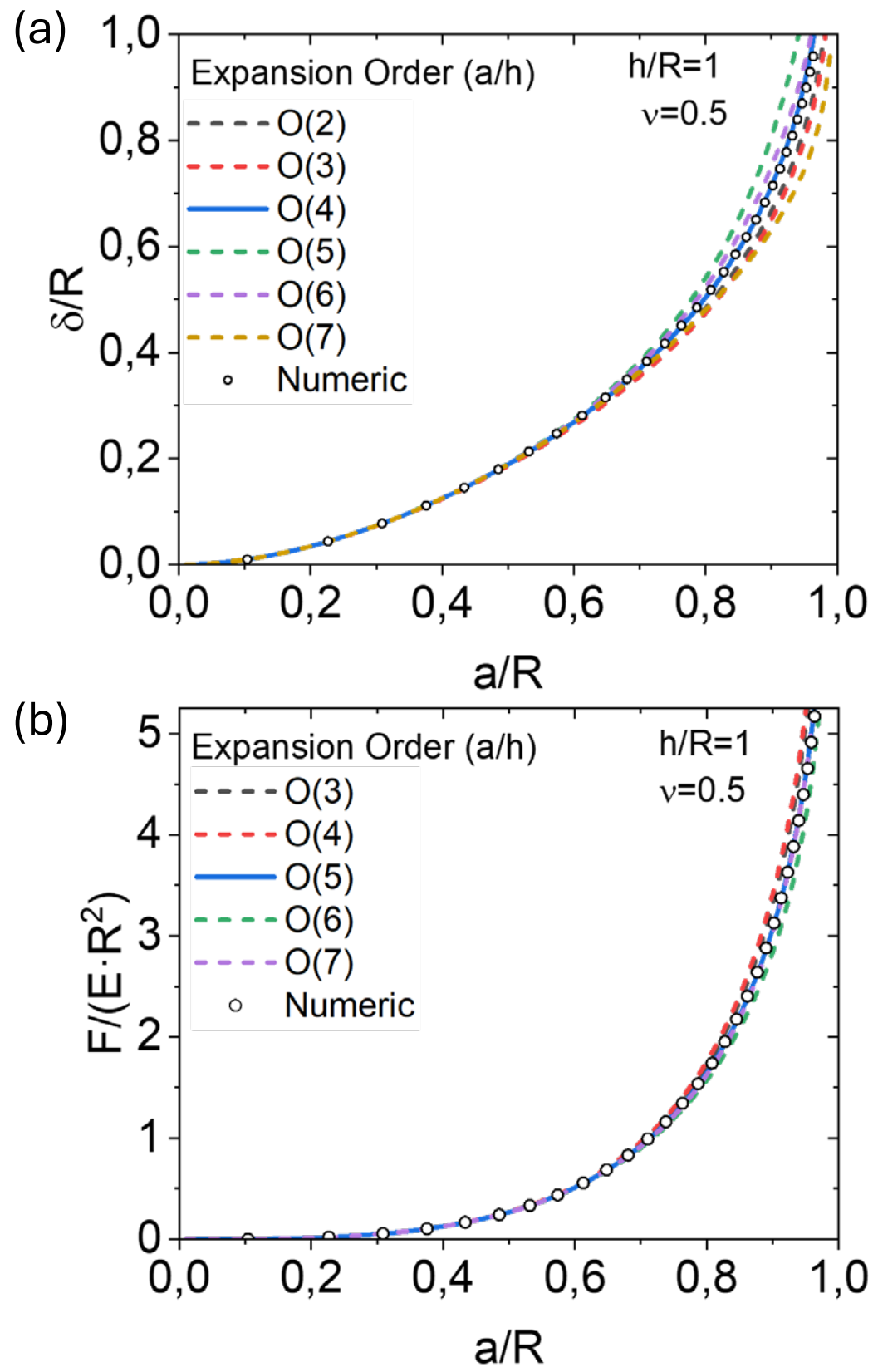

**Figure B1.** (a) (symbols) Normalised indentation versus contact area for a spherical indenter on a film with $h/R$=1, obtained numerically by solving the contact mechanics integral equation, Eq. (9). (Dashed lines) Indentation versus contact area curves predicted by the analytical series expansion solution of Dhaliwal and Rau [17] for a rigid substrate, Eq. (B1), evaluated at different orders in *a/h* (from O(2) to O(7)). The continuous line highlights the order (in this case O(4)) that best fits the numerical data in the whole range of indentations. (b) Idem but for the case of the force, by using the O(4) expression for the indentation in it. In this case, the best agreement is obtained by considering up to the O(5) term in *a/h* in the force series expansion in Eq. (B1).

Figure B1b shows the same study but for the normalised force $F/(E{\cdot}R^2)$ with the indentation kept fixed at O(4). This time, the expression up to O(5), provides the best agreement. Based on these results, the truncated analytical series expansion solution to order $n_{inf,F}$=5 in the force and to order $n_{inf,\delta}$=4 in the indentation provides the best description of both magnitudes in the whole range of indentations. This truncated analytical solution can be written explicitly as

$$\begin{aligned}
F=\frac{4aE}{1-\upsilon^2}\Bigg[&\delta\alpha_0^{(0)}+\left(\delta\alpha_0^{(1)}-\beta_0^{(1)}\right)\left(\frac{a}{h}\right)^1+\left(\delta\alpha_0^{(2)}-\beta_0^{(2)}\right)\left(\frac{a}{h}\right)^2+\left(\delta\alpha_0^{(3)}-\beta_0^{(3)}\right)\left(\frac{a}{h}\right)^3+\left(\delta\alpha_0^{(4)}-\beta_0^{(4)}\right)\left(\frac{a}{h}\right)^4\\
&+\left(\delta\alpha_0^{(5)}-\beta_0^{(5)}\right)\left(\frac{a}{h}\right)^5+\frac{1}{3}\left(\delta\alpha_1^{(3)}-\beta_1^{(3)}\right)\left(\frac{a}{h}\right)^3+\frac{1}{3}\left(\delta\alpha_1^{(4)}-\beta_1^{(4)}\right)\left(\frac{a}{h}\right)^4+\frac{1}{3}\left(\delta\alpha_1^{(5)}-\beta_1^{(5)}\right)\left(\frac{a}{h}\right)^5\\
&+\frac{1}{3}\left(\delta\alpha_2^{(5)}-\beta_2^{(5)}\right)\left(\frac{a}{h}\right)^5-\frac{1}{4}a\left\{\frac{2R}{a}+\left(1+\frac{R^2}{a^2}\right)\log\left(\frac{R+a}{R-a}\right)\right\}\Bigg]\\
\delta=&\frac{\left(\frac{a}{h}\right)^1\beta_0^{(1)}+\left(\frac{a}{h}\right)^2\beta_0^{(2)}+\left(\frac{a}{h}\right)^3\left(\beta_0^{(3)}+\beta_1^{(3)}\right)+\left(\frac{a}{h}\right)^4\left(\beta_0^{(4)}+\beta_1^{(4)}\right)+\frac{1}{2}a\log\left(\frac{R+a}{R-a}\right)}{\alpha_0^{(0)}+\left(\frac{a}{h}\right)\alpha_0^{(1)}+\left(\frac{a}{h}\right)^2\alpha_0^{(2)}+\left(\frac{a}{h}\right)^3\left(\alpha_0^{(3)}+\alpha_1^{(3)}\right)+\left(\frac{a}{h}\right)^4\left(\alpha_0^{(4)}+\alpha_1^{(4)}\right)}
\end{aligned}\tag{B5}$$

where the different parameters and functions appearing in Eq. (B5) must be evaluated numerically for each value of Poisson's ratio, *v*, by using Eq. (B3) (see the Mathematica 14 notebook provided as Supporting Information). For instance, for *v*=0.5 one has:

$$\begin{aligned}
&\alpha_0^{(0)}=1,\quad \alpha_0^{(1)}=1.12695,\quad \alpha_0^{(2)}=1.27003,\quad \alpha_0^{(3)}=1.02477,\\
&\alpha_0^{(4)}=0.238683,\quad \alpha_1^{(3)}=-1.21947,\quad \alpha_1^{(4)}=-1.37429
\end{aligned}\tag{B6}$$

and

$$\begin{aligned}
\beta_0^{(1)}&=0.281739a\left(\frac{2R}{a}+\left(1-\frac{R^2}{a^2}\right)\log\left(\frac{R+a}{R-a}\right)\right)\\
\beta_0^{(2)}&=0.317507a\left(\frac{2R}{a}+\left(1-\frac{R^2}{a^2}\right)\log\left(\frac{R+a}{R-a}\right)\right)\\
\beta_1^{(3)}&=-0.304867a\left(\frac{2R}{a}+\left(1-\frac{R^2}{a^2}\right)\log\left(\frac{R+a}{R-a}\right)\right)\\
\beta_1^{(4)}&=-0.334357a\left(\frac{2R}{a}+\left(1-\frac{R^2}{a^2}\right)\log\left(\frac{R+a}{R-a}\right)\right)\\
\beta_2^{(5)}&=0.185621a\left(\frac{2R}{a}+\left(1-\frac{R^2}{a^2}\right)\log\left(\frac{R+a}{R-a}\right)\right)
\end{aligned}\tag{B7}$$

$$\beta_0^{(3)} = 0.357816a\left(\frac{2R}{a} + \left(1 - \frac{R^2}{a^2}\right)\log\left(\frac{R+a}{R-a}\right)\right) - 0.152434a\left(2\left(\frac{R}{3a} + \frac{R^3}{a^3}\right) + \left(1 - \frac{R^4}{a^4}\right)\log\left(\frac{R+a}{R-a}\right)\right)$$

$$\beta_0^{(4)} = 0.174193a\left(\frac{2R}{a} + \left(1 - \frac{R^2}{a^2}\right)\log\left(\frac{R+a}{R-a}\right)\right) - 0.171785a\left(2\left(\frac{R}{3a} + \frac{R^3}{a^3}\right) + \left(1 - \frac{R^4}{a^4}\right)\log\left(\frac{R+a}{R-a}\right)\right)$$

$$\beta_1^{(5)} = -0.38719a\left(\frac{2R}{a} + \left(1 - \frac{R^2}{a^2}\right)\log\left(\frac{R+a}{R-a}\right)\right) + 0.556863a\left(2\left(\frac{R}{3a} + \frac{R^3}{a^3}\right) + \left(1 - \frac{R^4}{a^4}\right)\log\left(\frac{R+a}{R-a}\right)\right)$$

$$\beta_0^{(5)} = -0.0618205a\left(\frac{2R}{a} + \left(1 - \frac{R^2}{a^2}\right)\log\left(\frac{R+a}{R-a}\right)\right) - 0.193594a\left(2\left(\frac{R}{3a} + \frac{R^3}{a^3}\right) + \left(1 - \frac{R^4}{a^4}\right)\log\left(\frac{R+a}{R-a}\right)\right)$$
$$+0.0618736a\left[2\left(\frac{R}{5a} + \frac{R^3}{3a^3} + \frac{R^5}{a^5}\right) + \left(1 - \frac{R^6}{a^6}\right)\log\left(\frac{R+a}{R-a}\right)\right]$$

By substituting Eqs. (B6) and (B7) into Eq. (B5), we obtained

$$\begin{cases} F(a) = \frac{8Ea}{3}\Bigg[\left(1 + 1.12695\left(\frac{a}{h}\right) + 1.27003\left(\frac{a}{h}\right)^2 + 0.61828\left(\frac{a}{h}\right)^3 - 0.21941\left(\frac{a}{h}\right)^4 - 0.48779\left(\frac{a}{h}\right)^5\right)\delta(a) \\ \quad - \left(\frac{1}{4} + 0.281739\left(\frac{a}{h}\right) + 0.317507\left(\frac{a}{h}\right)^2 + 0.256194\left(\frac{a}{h}\right)^3 + 0.0596706\left(\frac{a}{h}\right)^4 - 0.153756\left(\frac{a}{h}\right)^5\right)a\left(\frac{2R}{a} + \left(1 - \frac{R^2}{a^2}\right)\log\left(\frac{R+a}{R-a}\right)\right) \\ \quad + \left(0.152434\left(\frac{a}{h}\right)^3 + 0.171786\left(\frac{a}{h}\right)^4 + 0.00797409\left(\frac{a}{h}\right)^5\right)a\left(2\left(\frac{R}{3a} + \frac{R^3}{a^3}\right) + \left(1 - \frac{R^4}{a^4}\right)\log\left(\frac{R+a}{R-a}\right)\right) \\ \quad - \left(0.0618736\left(\frac{a}{h}\right)^5\right)a\left(2\left(\frac{R}{5a} + \frac{R^3}{3a^3} + \frac{R^5}{a^5}\right) + \left(1 - \frac{R^6}{a^6}\right)\log\left(\frac{R+a}{R-a}\right)\right)\Bigg] & a/h < 1, \quad a/R < 1 \\ \delta(a) = \dfrac{\frac{1}{2}a\log\left(\frac{R+a}{R-a}\right) + a\left(\frac{2R}{a} + \left(1 - \frac{R^2}{a^2}\right)\log\left(\frac{R+a}{R-a}\right)\right)\left[0.281739\left(\frac{a}{h}\right) + 0.317507\left(\frac{a}{h}\right)^2 + 0.052943\left(\frac{a}{h}\right)^3 - 0.160164\left(\frac{a}{h}\right)^4\right]}{1 + 1.12695\left(\frac{a}{h}\right) + 1.27003\left(\frac{a}{h}\right)^2 - 0.194696\left(\frac{a}{h}\right)^3 - 1.1356\left(\frac{a}{h}\right)^4} \\ \quad + \dfrac{a\left(2\left(\frac{R}{3a} + \frac{R^3}{a^3}\right) + \left(1 - \frac{R^4}{a^4}\right)\log\left(\frac{R+a}{R-a}\right)\right)\left[-0.152434\left(\frac{a}{h}\right)^3 - 0.171785\left(\frac{a}{h}\right)^4\right]}{1 + 1.12695\left(\frac{a}{h}\right) + 1.27003\left(\frac{a}{h}\right)^2 - 0.194696\left(\frac{a}{h}\right)^3 - 1.1356\left(\frac{a}{h}\right)^4} \end{cases} \tag{B8}$$

Figure B2 (continuous colour lines) shows the normalised force and indentation curves predicted by Eq. (B8) as a function of contact radius for a wide range of *h/R* ratios. The agreement with the numerically calculated data is excellent for all thicknesses and indentations if the condition $a/h<1$ is satisfied. For other values of Poisson's ratio $\nu$, one can proceed similarly, and a similar agreement is obtained. We note that a truncated expression of the Dhaliwal and Rau solution considered in Ref. [24] for the case $h/R=2$ required a lower order. But for $h/R=1$ (and lower *h/R* values) higher order terms need to be considered.

A simpler, although less accurate, expression than that in Eq. (B8) can be obtained by developing the expressions in Eq. (8) as a true series of expansion in *a/h* up to O(4). For $\nu=0.5$ one simply obtains

$$
\begin{cases}
F=\dfrac{E}{\left(1-0.5^{2}\right)}\left(\left(R^{2}+a^{2}\right)\operatorname{arctanh}\left(\dfrac{a}{R}\right)-aR\right)\left[1+0.8555\left(\dfrac{a}{h}\right)^{3}\right] \\
\delta=a\operatorname{arctanh}\left(\dfrac{a}{R}\right)+a\left(\left(1+\left(\dfrac{R}{a}\right)^{2}\right)\operatorname{arctanh}\left(\dfrac{a}{R}\right)-\dfrac{R}{a}\right)\left[-0.5664\dfrac{a}{h}-0.4846\left(\dfrac{a}{h}\right)^{4}\right] \quad a\leq R,h \\
+0.1069a\left(\left(3\left(\dfrac{R}{a}\right)^{4}+6\left(\dfrac{R}{a}\right)^{2}+7\right)\operatorname{arctanh}\left(\dfrac{a}{R}\right)-\dfrac{R}{a}\left(3\left(\dfrac{R}{a}\right)^{2}+7\right)\right)\left(\dfrac{a}{h}\right)^{3}
\end{cases}
\tag{B9}
$$

Equation (B9) has the functional form predicted by Argatov [21] for the series expansion solutions of axisymmetric indenters expressed as series expansions in terms of *a/h* up to O(4) and provides reasonably good agreement with the exact numerical calculations (dashed lines in Figure B2) with relative errors below 6% in its range of validity, i.e., when *a/h*<1. We note that Eq. (B9) can also be derived by following the procedure described by Dimitriadis et al. [5] and Garcia et al. [16] based on the use of the reciprocal theorem and the expression for the pressure distribution of a flat punch, showing the equivalency between both approaches (see Supporting Information).

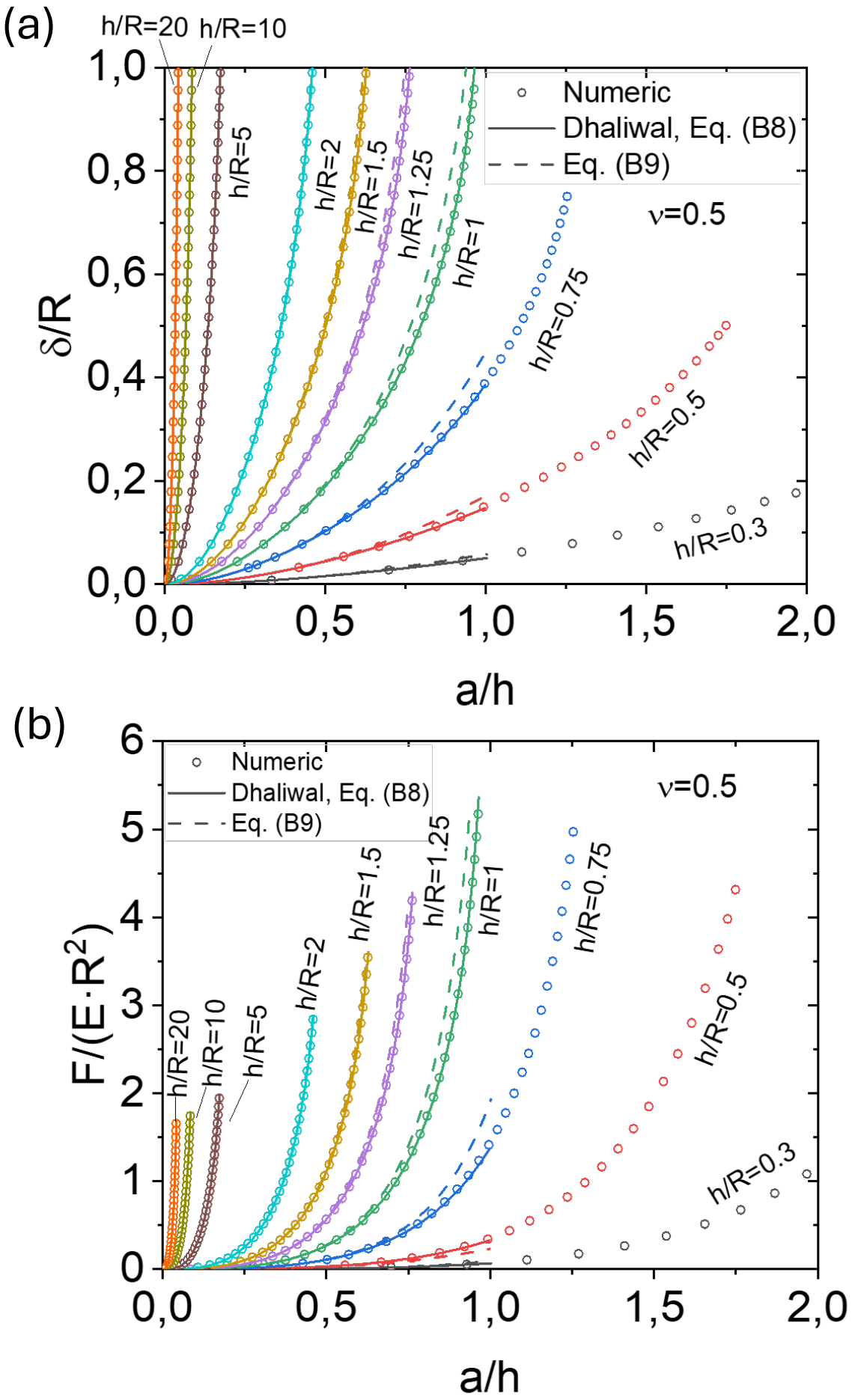

**Figure B2.** (a) and (b) (symbols) are normalised numerically calculated values of the force and indentation, respectively, versus indentation for different ratios of *h/R* ranging from 0.3 to 20,

calculated numerically by solving Eq. (8) with Eqs. (11) and (12) for $\nu$=0.5. (Continuous and dashed lines) idem but calculated using the analytical formulae in Eqs. (B8) and (B9), respectively. Note that the contact radius has been normalised by the thickness of these plots to help visualise the different curves and data.

**Appendix C: Functional dependency of the explicit analytical sphere model (Eq. (4)).**

The functional dependence of the sphere model in Eq. (4) has been obtained by first performing a Taylor's series expansion of the truncated Dhaliwal and Raus analytical solution for the indentation in Eq. (B8) in terms of *a/h* up to O(4), followed by an inversion of the series to obtain the contact radius *a* as a series expansion of the indentation, $\delta$. For $\nu$=0.5 we obtained

$$a \approx (R\delta)^{1/2}\left[1-0.166666\frac{\delta}{R}+0.375652\left(\frac{R\delta}{h^2}\right)^{1/2}+0.352785\left(\frac{R\delta}{h^2}\right)\right] \quad \text{(C1)}$$

These calculations were done using the symbolic calculation capabilities of Mathematica 14.2 from Wolfram. Then, we substituted Eq. (C1) into the force expression of the truncated Dhaliwal and Rau's analytical solution in Eq. (B8), normalized it by the explicit version of Sneddon's formula (Eq. (2)) and performed a Taylor's series expansion up to terms of order O(5/2) in $\delta$, giving rise to Eq. (4), with the following values of the parameters for $\nu$=0.5

$$\begin{aligned} &\nu = 0.5 \quad \text{(non-fitted)} \\ &a_1 = 1.12695; \quad a_2 = 1.48170; \quad a_3 = 0.48835; \\ &a_4 = 1.66113; \quad a_5 = 0.99485; \quad a_6 = 2.35509; \\ &a_7 = 0.256785; \; a_8 = 0.81804; \quad a_9 = 1.98529; \end{aligned} \quad \text{(C2)}$$

where a coefficient for the first term in $(\delta/R)^2$ was set to zero, since its very small value did not improve the overall accuracy of the expression. When Eq. (4) is evaluated with the values of the parameters given in Eq. (C2), the agreement with the numerical data is only good in the small indentation range due to the successive approximations made to arrive at it. If instead the coefficients in Eq. (4) are treated as phenomenological parameters to be determined by a least squares fit of Eq. (4) to the data calculated numerically, one obtains almost perfect agreement with the numerical data. For the case that $\nu$=0.5 and the numerical data in Figure 3, we obtained

$$\begin{aligned}&\nu = 0.5 \quad \text{(Fitted)}\\ &a_1 = 1.01345; \quad a_2 = 2.24003; \quad a_3 = -0.470293;\\ &a_4 = 0.93412; \quad a_5 = -0.455595; \quad a_6 = -0.124456;\\ &a_7 = 0.20410; \; a_8 = -0.938828; \quad a_9 = 0.0291373;\end{aligned} \tag{C3}$$

leading to Eq. (5) of the main text, which shows a relative error of less than 1.3% with the numerically calculated data. The least squares fits were performed by using Mathematica 14.2 from Wolfram. A notebook is provided as Supporting Information. A similar procedure can be followed for other values of the Poisson's ratio. For instance, for $\nu$=0.25 by using the numerical data shown in Figure C1a we obtained for the $a_n$ coefficients

$$\begin{aligned}&\nu = 0.25 \quad \text{(Fitted)}\\ &a_1 = 0.76318; \quad a_2 = 1.4062; \quad a_3 = -0.161757;\\ &a_4 = -0.412556; \quad a_5 = -0.49387; \quad a_6 = 0.0158609;\\ &a_7 = 0.0276841; \; a_8 = 0.131375; \quad a_9 = 0.007778;\end{aligned} \tag{C4}$$

The expression obtained shows an excellent agreement with the numerically calculated data for this value of Poisson's ratio in the whole indentation and thickness ranges (see Figure C1a), with a relative error, in this case, smaller than 0.5%. The values of the coefficients for other Poisson's ratios are given in Table C1 and are plotted in Figure C1b. Values of the Poisson's ratio not explicitly calculated can be obtained by mathematical interpolation, since no simple phenomenological expressions could be found to describe all of them analytically.

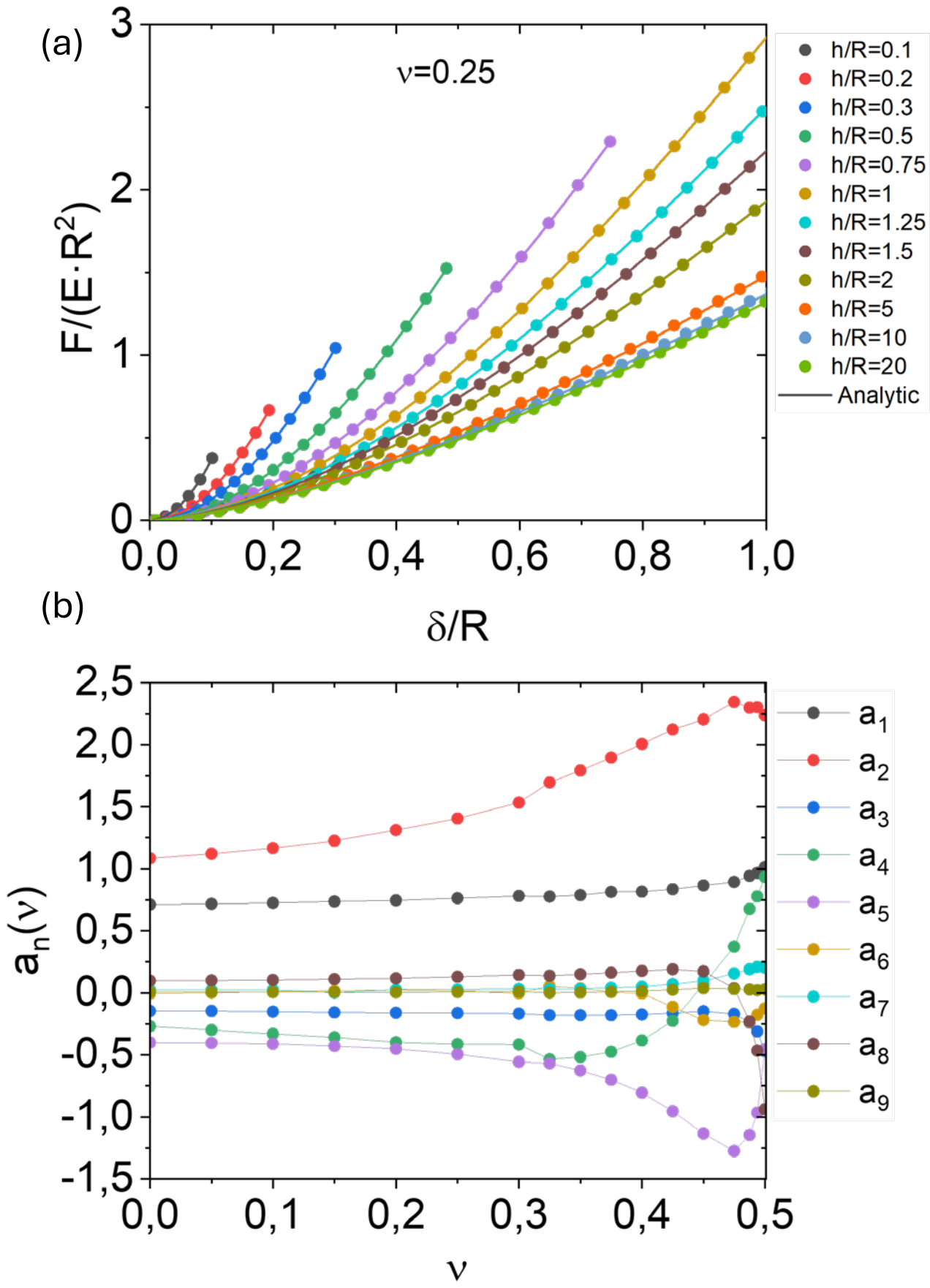

**Figure C1.** (a) (Symbols) Numerically calculated force indentation curves for a spherical indenter on a thin film for different thicknesses for a Poisson's ratio $\nu$=0.25. (continuous lines) Prediction of Eq. (4) with the coefficients determined for $\nu$=0.25 (Eq. (C4)). (b) Dependence of the coefficients $a_n(\nu)$ on the Poisson's ratio ν in Eq. (4) (see the actual values in Table C1).

**Table C1.** Values of the phenomenological coefficients $a_n$ in Eq. (4) as function of Poisson's ratio, $\nu$

| **ν** | **$a_1$** | **$a_2$** | **$a_3$** | **$a_4$** | **$a_5$** | **$a_6$** | **$a_7$** | **$a_8$** | **$a_9$** |
|---|---|---|---|---|---|---|---|---|---|
| **0** | 0.71304 | 1.08600 | -0.14552 | -0.26861 | -0.39948 | -0.00599 | 0.02439 | 0.09881 | 0.00793 |
| **0.05** | 0.71891 | 1.12237 | -0.14552 | -0.29872 | -0.40294 | 0.00307 | 0.0236 | 0.10117 | 0.00698 |
| **0.1** | 0.72685 | 1.16852 | -0.15089 | -0.32954 | -0.40985 | 0.01104 | 0.02434 | 0.10456 | 0.00627 |
| **0.15** | 0.73721 | 1.22667 | -0.15648 | -0.35824 | -0.42847 | 0.01582 | 0.00281 | 0.11217 | 0.00616 |
| **0.2** | 0.74653 | 1.31300 | -0.15942 | -0.39813 | -0.45057 | 0.02336 | 0.02564 | 0.11851 | 0.00575 |
| **0.25** | 0.76318 | 1.40620 | -0.16176 | -0.41256 | -0.49387 | 0.01586 | 0.02768 | 0.13138 | 0.00778 |
| **0.3** | 0.78221 | 1.53432 | -0.16643 | -0.41544 | -0.5558 | -0.00118 | 0.03231 | 0.14607 | 0.01141 |
| **0.325** | 0.77724 | 1.69806 | -0.17888 | -0.53416 | -0.56704 | 0.05464 | 0.02976 | 0.13947 | 0.0016 |
| **0.35** | 0.78826 | 1.79348 | -0.17869 | -0.51763 | -0.62498 | 0.03696 | 0.03332 | 0.15109 | 0.00449 |
| **0.375** | 0.81478 | 1.89819 | -0.17872 | -0.47411 | -0.70118 | 0.00725 | 0.04122 | 0.16401 | 0.00900 |
| **0.4** | 0.81733 | 2.00782 | -0.17296 | -0.38316 | -0.80438 | -0.00416 | 0.05053 | 0.17686 | 0.01597 |
| **0.425** | 0.83612 | 2.12370 | -0.16217 | -0.22606 | -0.95464 | -0.11203 | 0.07075 | 0.19141 | 0.02526 |
| **0.45** | 0.86651 | 2.20562 | -0.14794 | 0.06980 | -1.13286 | -0.21878 | 0.09861 | 0.17420 | 0.03804 |
| **0.475** | 0.89505 | 2.34625 | -0.16850 | 0.37311 | -1.27422 | -0.23400 | 0.15742 | 0.03864 | 0.03156 |
| **0.4875** | 0.94369 | 2.29927 | -0.23759 | 0.67819 | -1.14440 | -0.23456 | 0.19097 | -0.23003 | 0.02857 |
| **0.49375** | 0.96771 | 2.30345 | -0.31014 | 0.77793 | -0.96550 | -0.17738 | 0.20973 | -0.46535 | 0.02271 |
| **0.5** | 1.01345 | 2.24003 | -0.47029 | 0.93412 | -0.45560 | -0.12446 | 0.2041 | -0.93883 | 0.02914 |

## Acknowledgments

This work received funding from the European Union's Horizon 2020 programme under the EICPathfinder PRINGLE project (grant agreement No. 101046719); the Horizon Europe programme under the MSCA DN SPM4.0 project (grant agreement No. 101168976); from the Ministerio de Ciencia e Innovación through grant no. PID2022-142297NB-I00 (BIOMEDSPM40); from the Generalitat de Catalunya through CERCA; and from the ICREA foundation (ICREA Academia award to G.G.). BC acknowledges support from the Ministerio de Ciencia, Innovación y Universidades (FPI grant, PREP2022-000801). AC acknowledges funding from the Fondazione Bartolo Longo and Academist/Sysmex Diagnostics SE. We would also like to acknowledge the European Research Council Grant No. 850936 (PANDORA); the Fondazione Cariplo (Grant No. 2019-4278); and the Ministry of Education, Universities and Research (through the PRIN program, Grant No. 20205B2HZE). We also thank the H2020-ERC-2018-CoG CheSSTaG (769798) and the Spanish Ministry of Science and Innovation, Projects: ID+IPID2020-119914RB-I00, I+D+I EQC2019-005937-P, and I+D+I PID2023-149206OB-I00.

**Data Availability Statement**: All data generated or analysed during this study are included in this document and in the supporting information files.

**References**

[1] M. Krieg *et al.*, “Atomic force microscopy-based mechanobiology,” *Nature Reviews Physics*, vol. 1, no. 1, pp. 41–57, Nov. 2018, doi: 10.1038/s42254-018-0001-7.

[2] P. D. Garcia, C. R. Guerrero, and R. Garcia, “Nanorheology of living cells measured by AFM-based force–distance curves,” *Nanoscale*, vol. 12, no. 16, pp. 9133–9143, Apr. 2020, doi: 10.1039/C9NR10316C.

[3] A. Viljoen *et al.*, “Force spectroscopy of single cells using atomic force microscopy,” *Nature Reviews Methods Primers*, vol. 1, no. 1, p. 63, Sep. 2021, doi: 10.1038/s43586-021-00062-x.

[4] R. Garcia and J. R. Tejedor, “Advances in nanomechanical property mapping by atomic force microscopy,” *Nanoscale Adv.*, vol. 7, no. 20, pp. 6286–6307, Oct. 2025, doi: 10.1039/D5NA00702J.

[5] E. K. Dimitriadis, F. Horkay, J. Maresca, B. Kachar, and R. S. Chadwick, “Determination of Elastic Moduli of Thin Layers of Soft Material Using the Atomic Force Microscope,” *Biophys. J.*, vol. 82, no. 5, pp. 2798–2810, May 2002, doi: 10.1016/S0006-3495(02)75620-8.

[6] R. E. Mahaffy, S. Park, E. Gerde, J. Käs, and C. K. Shih, “Quantitative Analysis of the Viscoelastic Properties of Thin Regions of Fibroblasts Using Atomic Force Microscopy,” *Biophys. J.*, vol. 86, no. 3, pp. 1777–1793, Mar. 2004, doi: 10.1016/S0006-3495(04)74245-9.

[7] P. Carl and H. Schillers, “Elasticity measurement of living cells with an atomic force microscope: data acquisition and processing,” *Pflugers Arch.*, vol. 457, no. 2, pp. 551–559, Nov. 2008, doi: 10.1007/s00424-008-0524-3.

[8] A. R. Harris and G. T. Charras, “Experimental validation of atomic force microscopy-based cell elasticity measurements,” *Nanotechnology*, vol. 22, no. 34, p. 345102, Aug. 2011, doi: 10.1088/0957-4484/22/34/345102.

[9] L. Puricelli, M. Galluzzi, C. Schulte, A. Podestà, and P. Milani, “Nanomechanical and topographical imaging of living cells by atomic force microscopy with colloidal probes,” *Review of Scientific Instruments*, vol. 86, no. 3, Mar. 2015, doi: 10.1063/1.4915896.

[10] H. Schillers *et al.*, “Standardized Nanomechanical Atomic Force Microscopy Procedure (SNAP) for Measuring Soft and Biological Samples,” *Sci. Rep.*, vol. 7, no. 1, p. 5117, Jul. 2017, doi: 10.1038/s41598-017-05383-0.

[11] M. Chighizola, L. Puricelli, L. Bellon, and A. Podestà, "Large colloidal probes for atomic force microscopy: Fabrication and calibration issues," *Journal of Molecular Recognition*, vol. 34, no. 1, Jan. 2021, doi: 10.1002/jmr.2879.

[12] H. Plank *et al.*, "Focused Electron Beam-Based 3D Nanoprinting for Scanning Probe Microscopy: A Review," *Micromachines (Basel).*, vol. 11, no. 1, p. 48, Dec. 2019, doi: 10.3390/mi11010048.

[13] C. M. Segedin, "The relation between load and penetration for a spherical punch," *Mathematika*, vol. 4, no. 2, pp. 156–161, Dec. 1957, doi: 10.1112/S0025579300001236.

[14] I. N. Sneddon, "The relation between load and penetration in the axisymmetric boussinesq problem for a punch of arbitrary profile," *Int. J. Eng. Sci.*, vol. 3, no. 1, pp. 47–57, May 1965, doi: 10.1016/0020-7225(65)90019-4.

[15] E. Popova and V. L. Popov, "The History of 'Sneddon's' solution in contact mechanics," *PAMM*, vol. 21, no. 1, Dec. 2021, doi: 10.1002/pamm.202100048.

[16] P. D. Garcia and R. Garcia, "Determination of the Elastic Moduli of a Single Cell Cultured on a Rigid Support by Force Microscopy," *Biophys. J.*, vol. 114, no. 12, pp. 2923–2932, Jun. 2018, doi: 10.1016/j.bpj.2018.05.012.

[17] J. Domke and M. Radmacher, "Measuring the Elastic Properties of Thin Polymer Films with the Atomic Force Microscope," *Langmuir*, vol. 14, no. 12, pp. 3320–3325, Jun. 1998, doi: 10.1021/la9713006.

[18] B. B. Akhremitchev and G. C. Walker, "Finite Sample Thickness Effects on Elasticity Determination Using Atomic Force Microscopy," *Langmuir*, vol. 15, no. 17, pp. 5630–5634, Aug. 1999, doi: 10.1021/la980585z.

[19] N. Gavara and R. S. Chadwick, "Determination of the elastic moduli of thin samples and adherent cells using conical atomic force microscope tips," *Nat. Nanotechnol.*, vol. 7, no. 11, pp. 733–736, Nov. 2012, doi: 10.1038/nnano.2012.163.

[20] P. Hermanowicz, "Determination of Young's modulus of samples of arbitrary thickness from force distance curves: numerical investigations and simple approximate formulae," *Int. J. Mech. Sci.*, vol. 193, p. 106138, Mar. 2021, doi: 10.1016/j.ijmecsci.2020.106138.

[21] I. Argatov and X. Jin, "Self-consistent approximations for the frictionless paraboloidal and conical depth-sensing indentation: The generalized bottom effect," *Int. J. Solids Struct.*, vol. 291, p. 112713, Apr. 2024, doi: 10.1016/j.ijsolstr.2024.112713.

[22] R. S. Dhaliwal and I. S. Rau, "The axisymmetric boussinesq problem for a thick elastic layer under a punch of arbitrary profile," *Int. J. Eng. Sci.*, vol. 8, no. 10, pp. 843–856, Oct. 1970, doi: 10.1016/0020-7225(70)90086-8.

[23] R. Long, M. S. Hall, M. Wu, and C.-Y. Hui, "Effects of Gel Thickness on Microscopic Indentation Measurements of Gel Modulus," *Biophys. J.*, vol. 101, no. 3, pp. 643–650, Aug. 2011, doi: 10.1016/j.bpj.2011.06.049.

[24] B. L. Doss, K. Rahmani Eliato, K. Lin, and R. Ros, "Quantitative mechanical analysis of indentations on layered, soft elastic materials," *Soft Matter*, vol. 15, no. 8, pp. 1776–1784, 2019, doi: 10.1039/C8SM02121J.

[25] L. Dal Fabbro, H. Holuigue, M. Chighizola, and A. Podestà, "Validation of contact mechanics models for atomic force microscopy *via* finite elements analysis and nanoindentation experiments," *Nanoscale*, vol. 18, no. 19, pp. 10420–10436, Jun. 2026, doi: 10.1039/D5NR05344G.

[26] C. Valero, B. Navarro, D. Navajas, and J. M. García-Aznar, "Finite element simulation for the mechanical characterization of soft biological materials by atomic force microscopy," *J. Mech. Behav. Biomed. Mater.*, vol. 62, pp. 222–235, Sep. 2016, doi: 10.1016/j.jmbbm.2016.05.006.

[27] A. Perriot and E. Barthel, “Elastic contact to a coated half-space: Effective elastic modulus and real penetration,” *J. Mater. Res.*, vol. 19, no. 2, pp. 600–608, Feb. 2004, doi: 10.1557/jmr.2004.19.2.600.

[28] P. Müller, S. Abuhattum, S. Möllmert, E. Ulbricht, A. V. Taubenberger, and J. Guck, “nanite: using machine learning to assess the quality of atomic force microscopy-enabled nano-indentation data,” *BMC Bioinformatics*, vol. 20, no. 1, p. 465, Dec. 2019, doi: 10.1186/s12859-019-3010-3.

[29] R. Garcia, “Nanomechanical mapping of soft materials with the atomic force microscope: methods, theory and applications,” *Chem. Soc. Rev.*, vol. 49, no. 16, pp. 5850–5884, Aug. 2020, doi: 10.1039/D0CS00318B.

[30] Y.-C. Lin *et al.*, “Mechanosensing of substrate thickness,” *Phys. Rev. E*, vol. 82, no. 4, p. 041918, Oct. 2010, doi: 10.1103/PhysRevE.82.041918.

[31] C. R. Kuo, J. Xian, J. D. Brenton, K. Franze, and E. Sivaniah, “Complex Stiffness Gradient Substrates for Studying Mechanotactic Cell Migration,” *Advanced Materials*, vol. 24, no. 45, pp. 6059–6064, Nov. 2012, doi: 10.1002/adma.201202520.

[32] C. G. Tusan *et al.*, “Collective Cell Behavior in Mechanosensing of Substrate Thickness,” *Biophys. J.*, vol. 114, no. 11, pp. 2743–2755, Jun. 2018, doi: 10.1016/j.bpj.2018.03.037.

[33] Y. Liu, S. Nemec, C. Kopecky, M. H. Stenzel, and K. A. Kilian, “Hydrogel Microtumor Arrays to Evaluate Nanotherapeutics,” *Adv. Healthc. Mater.*, vol. 12, no. 14, Jun. 2023, doi: 10.1002/adhm.202201696.

[34] N. R. Richbourg *et al.*, “Precise control of synthetic hydrogel network structure via linear, independent synthesis-swelling relationships,” 2021. [Online]. Available: https://www.science.org

[35] N. Richbourg, M. E. Wechsler, J. J. Rodriguez-Cruz, and N. A. Peppas, "Model-based modular hydrogel design," *Nature Reviews Bioengineering*, vol. 2, no. 7, pp. 575–587, Mar. 2024, doi: 10.1038/s44222-024-00167-4.

[36] A. M. Kloxin, C. J. Kloxin, C. N. Bowman, and K. S. Anseth, "Mechanical Properties of Cellularly Responsive Hydrogels and Their Experimental Determination," *Advanced Materials*, vol. 22, no. 31, pp. 3484–3494, Aug. 2010, doi: 10.1002/adma.200904179.

[37] D. Lee, H. Zhang, and S. Ryu, "Elastic Modulus Measurement of Hydrogels," in *Cellulose-Based Superabsorbent Hydrogels. Polymers and Polymeric Composites: A Reference Series*, M. Mondal, Ed., Springer, Cham, 2018, pp. 1–21. doi: 10.1007/978-3-319-76573-0_60-1.

[38] A. Gandin *et al.*, "Simple yet effective methods to probe hydrogel stiffness for mechanobiology," *Sci. Rep.*, vol. 11, no. 1, p. 22668, Nov. 2021, doi: 10.1038/s41598-021-01036-5.

[39] N. R. Richbourg, M. K. Rausch, and N. A. Peppas, "Cross-evaluation of stiffness measurement methods for hydrogels," *Polymer (Guildf).*, vol. 258, p. 125316, Oct. 2022, doi: 10.1016/j.polymer.2022.125316.

[40] R. S. Fischer, K. A. Myers, M. L. Gardel, and C. M. Waterman, "Stiffness-controlled three-dimensional extracellular matrices for high-resolution imaging of cell behavior," *Nat. Protoc.*, vol. 7, no. 11, pp. 2056–2066, Nov. 2012, doi: 10.1038/nprot.2012.127.

[41] A. Elosegui-Artola *et al.*, "Rigidity sensing and adaptation through regulation of integrin types," *Nat. Mater.*, vol. 13, no. 6, pp. 631–637, Jun. 2014, doi: 10.1038/nmat3960.

[42] Z. Huo *et al.*, “Biomechanics of Macrophages on Disordered Surface Nanotopography,” *ACS Appl. Mater. Interfaces*, vol. 16, no. 21, pp. 27164–27176, May 2024, doi: 10.1021/acsami.4c04330.

[43] G. Zhou, B. Zhang, L. Wei, H. Zhang, M. Galluzzi, and J. Li, “Spatially Resolved Correlation between Stiffness Increase and Actin Aggregation around Nanofibers Internalized in Living Macrophages,” *Materials*, vol. 13, no. 14, p. 3235, Jul. 2020, doi: 10.3390/ma13143235.

[44] J. H. Wen *et al.*, “Interplay of matrix stiffness and protein tethering in stem cell differentiation,” *Nat. Mater.*, vol. 13, no. 10, pp. 979–987, Oct. 2014, doi: 10.1038/nmat4051.

[45] J. Comelles, V. Fernández-Majada, N. Berlanga-Navarro, V. Acevedo, K. Paszkowska, and E. Martínez, “Microfabrication of poly(acrylamide) hydrogels with independently controlled topography and stiffness,” *Biofabrication*, vol. 12, no. 2, p. 025023, Apr. 2020, doi: 10.1088/1758-5090/ab7552.

[46] R. Subramani *et al.*, “The Influence of Swelling on Elastic Properties of Polyacrylamide Hydrogels,” *Front. Mater.*, vol. 7, Jul. 2020, doi: 10.3389/fmats.2020.00212.

[47] H. Wu, L. Zhang, B. Zhao, W. Yang, and M. Galluzzi, “Deep learning strategy for small dataset from atomic force microscopy mechano-imaging on macrophages phenotypes,” *Front. Bioeng. Biotechnol.*, vol. 11, Oct. 2023, doi: 10.3389/fbioe.2023.1259979.

[48] C. Pérez-González *et al.*, “Mechanical compartmentalization of the intestinal organoid enables crypt folding and collective cell migration,” *Nat. Cell Biol.*, vol. 23, no. 7, pp. 745–757, Jul. 2021, doi: 10.1038/s41556-021-00699-6.

[49] E. Keogh, S. Chu, D. Hart, and M. Pazzani, “Segmenting time series: a survey and novel approach,” in *Data Mining in Time Series Databases*, World Scientific, 2004, pp. 1–21. doi: 10.1142/9789812565402_0001.

# Supporting Information

**Mechanical mapping of thin elastic films and living cells with spherical tip atomic force microscopy probes at large indentations**

*Gabriel Gomila[*,1,2], Mauricio Cano[1,2], Beatriz Cantero[1,2], Lara Aiassa[1], Loris Rizzello,[3,4] Giuseppe Battaglia,[1] Eleni Dalaka[1], Jordi Comelles[1], Annalisa Calò[*1,2]*

*[1]Institut de Bioenginyeria de Catalunya (IBEC), The Barcelona Institute of Science and Technology, c/ Baldiri i Reixac 11-15, 08028, Barcelona, Spain.*

*[2]Departament d'Enginyeria Electrònica i Biomèdica, Universitat de Barcelona, C/ Martí i Franquès 1, 08028, Barcelona, Spain*

*[3]National Institute of Molecular Genetics (INGM), Via Francesco Sforza, 35 20122 Milano, Italy.*

*[4]Department of Pharmaceutical Sciences, University of Milano, Via Luigi Mangiagalli, 25 20133 Milano, Italy.*

*Corresponding Authors: ggomila@ibecbarcelona.eu, acalo@ub.edu

**S1. Description of the Mathematica files (file: *Mathematica.zip*)**

*Mechanical_integral_Sphere_Paper_Version.nb*
Solves numerically the contact mechanics integral equation for a sphere indenter on a thin film (Eq. (8) with Eq. (11) subject to the boundary condition Eq. (12)) and performs least squares fit of the analytical model in Eq. (4) to determine the values of the phenomenological coefficients $a_n$.

*Mechanical_integral_eq_Paraboloid_Paper_Version.nb*
Solves numerically the contact mechanics integral equation for a paraboloid indenter on a thin film (Eq. (8) with Eq. (11) subject to the boundary condition Eq. (12)) and compares the results with existing analytical models Eqs. (A1)-(A3). It includes the calculated data for $\nu=0.5$.

*Dhaliwal_Rau_Sphere_analytic.nb*
Calculates the truncated Dhaliwal and Rau analytical solution for a sphere (Eqs. (B1)-(B4)) at the desired truncation orders and provides the corresponding analytical expressions.

**S2. Description of the Data files**

*Simulations.zip*
It contains the numerically calculated data obtained by solving the contact mechanics equation for a spherical indenter on a thin film (Eq. (8) with Eq. (11) subject to the boundary condition in Eq. (12)) for a range of values of the *h/R* and Poisson's ratios. For each Poisson's ratio, different values of *h/R* are considered, and for each of them, several values of the contact radius are calculated. To see the organisation of the data in the files, refer to the Export line in *Mechanical_integral_Sphere_Paper_Version.nb*.

*Repository.zip*
It contains the raw and calibrated experimental data files together with the fits of the theoretical models and the parameter maps obtained. Data of all figures in txt format are also included in this folder.
Note: Repository Data available upon request at https://doi.org/10.5281/zenodo.20765255.

## S3. Additional data for Fig. 5.

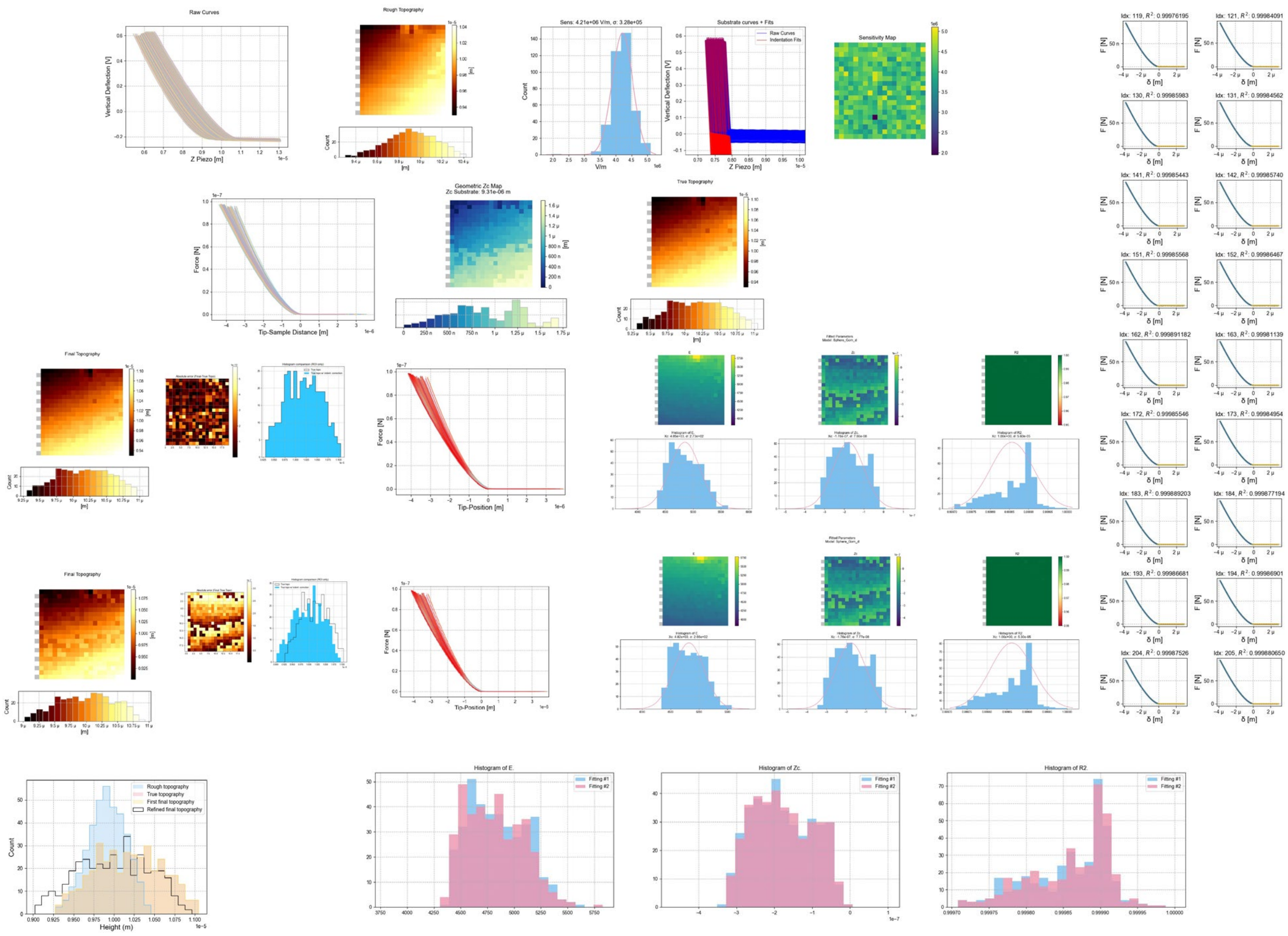

**Figure S1.** Additional data corresponding to the calibration and quantification of the force map in Fig. 5. The ensemble of images and data files resulting from the analysis is available in the data file Repository.rar, along with a description of the different data files and figures in the file Inventory.docx. Data processed with Mechanotool.

## S4. Hydrogel PAA film with uniform thickness

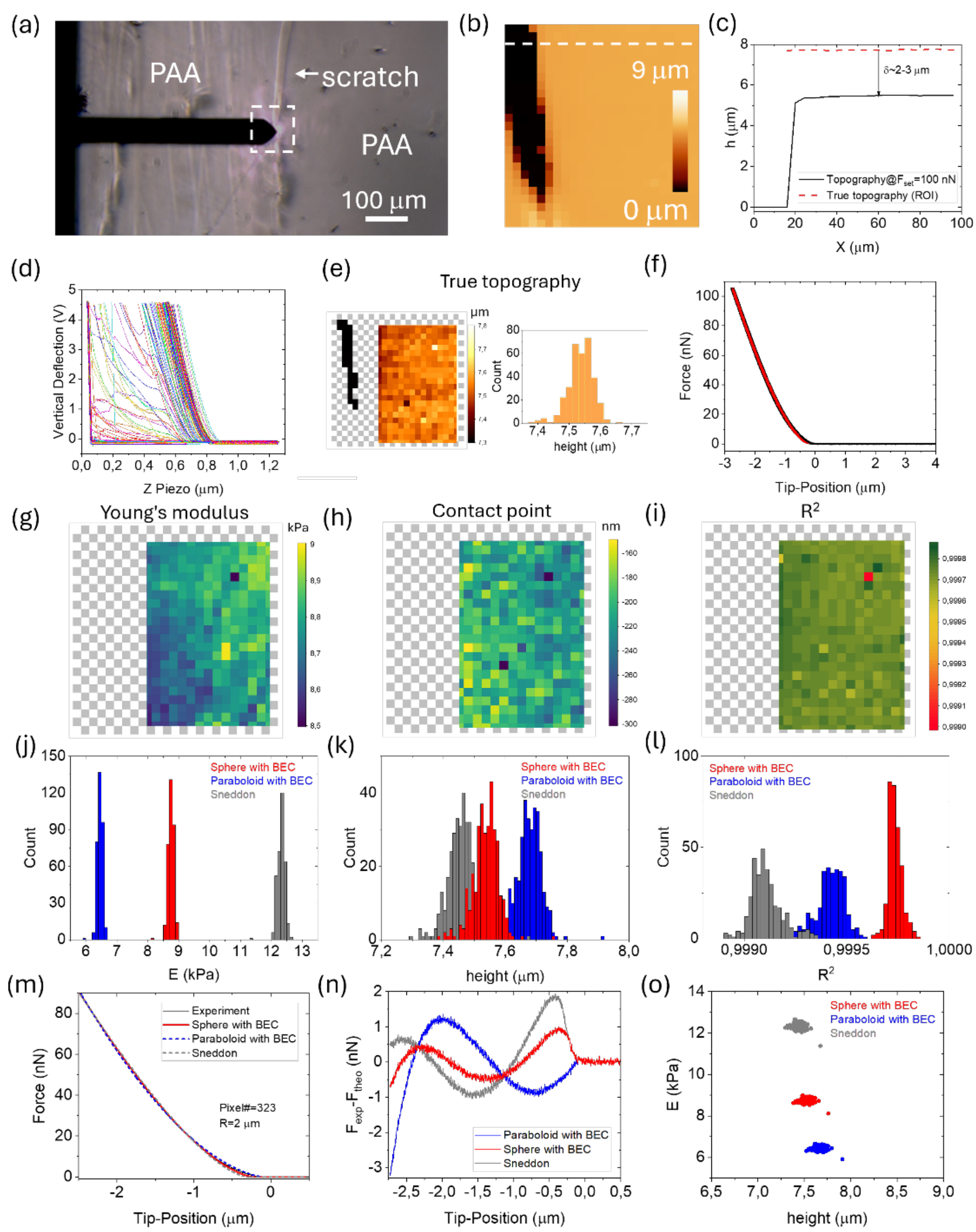

**Figure S2.** (a) Optical microscopy image of the AFM probe located close to the scratch performed on the PAA thin film. The highlighted dashed square represents the area of 100x100 μm$^2$ that was probed by the AFM. (b) Topography image obtained from the curve map at a set point of $F_{set}$=100 nN. (c) (continuous black line) Cross-sectional topographic profile along the white dashed line in (b). (red dashed line) True topography cross-section profile in the ROI. Deformations of the sample surface in the range 2-3 μm are obtained at the force set-point. (d) Raw deflection-piezo extension 25x25 curve map acquired in an area of 100x100 μm$^2$. (e) True topography of the ROI obtained after the data analysis, together with the height distribution histogram. (f) Calibrated force-indentation curves (black lines), together with the curves fitted with the sphere model proposed in the present work (red lines). (g) and (h) and (i) Maps of the Young's modulus, contact point variation and $R^2$ of the fits, respectively, obtained from the fitted curves in (f). (j), (k) and (l) Histograms of the extracted Young's modulus, true height and $R^2$ of the fits obtained from (g), (h) and (i), respectively. (m) (grey and red lines) Experimental and fitted force-indentation curve corresponding to a single pixel. For comparison, we also show the fitted curves obtained by using Sneddon's and the paraboloid with BEC models (grey and blue dashed lines). (n) Absolute error of the fitted curves in (m) for the different models as a function of indentation. (o) Extracted values of the Young's modulus as a function of the extracted true topography for the different models considered. Experimental parameters: biosphere B2000-CONT probe with k=0.49 N/m, R=2 μm, and approach velocity v=100 μm/s, $Z_{length}$=8 μm, pixel time 180 ms, points per curve 8000.

## S5. Additional data to Fig. 7

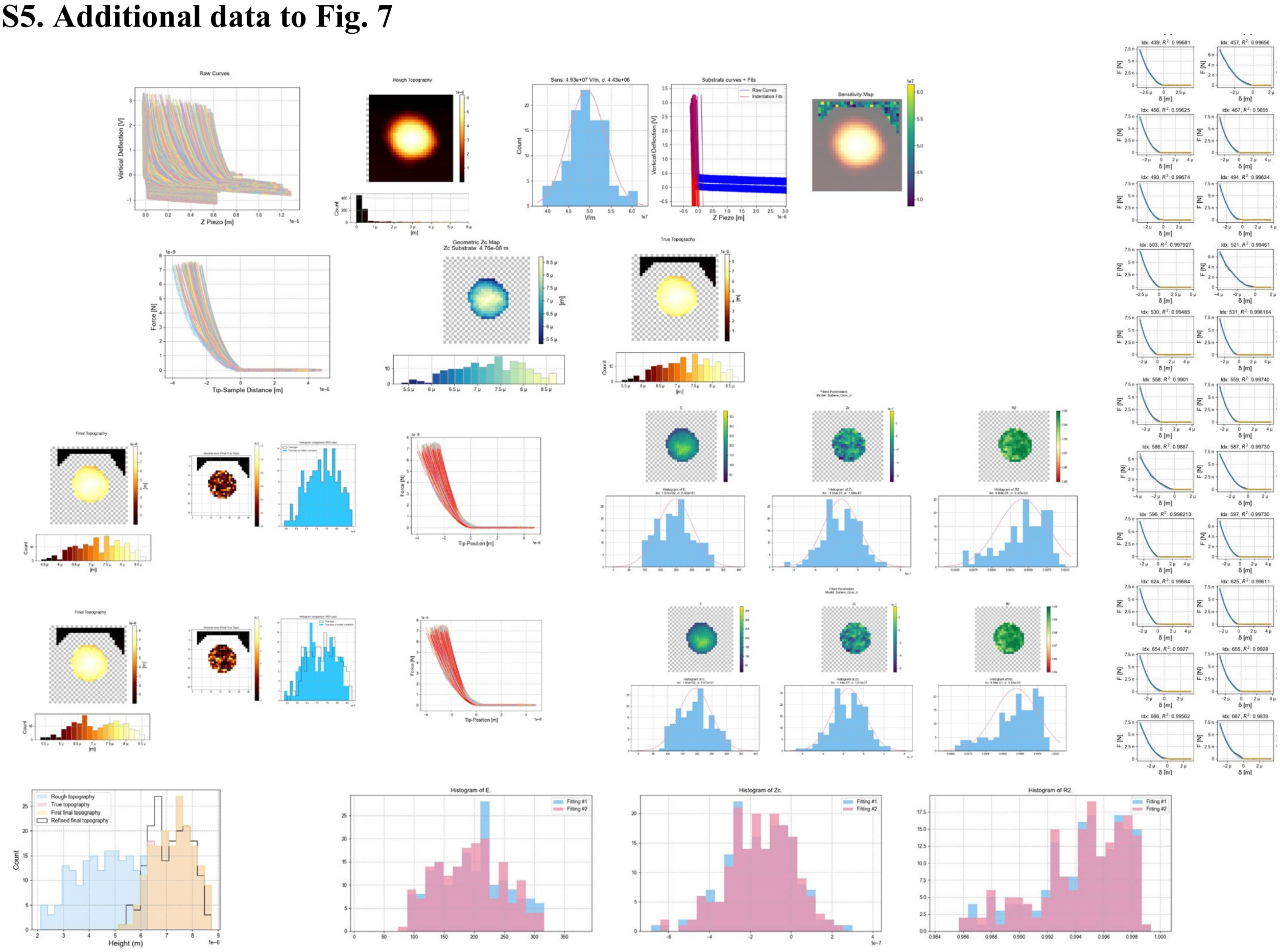

**Figure S3.** Additional data corresponding to the calibration and quantification of the force map in Fig. 7. The ensemble of images and data files resulting from the analysis are available in the data file Repository.rar, together with a description of the different data files and figures (in the file Inventory.docx). Data was processed with Mechanotool.

**S6. Derivation of Eq. (B9) by using the reciprocal theorem and the flat punch cylindrical series expansion solution.**

Equation (B9) can also be obtained independently by following the procedure described by Dimitriadis et al. [1] and Garcia et al. [2] based on the use of the reciprocal theorem and the expression for the pressure distribution of a flat punch, showing the equivalency of both approaches. According to the reciprocal theorem one can obtain the force due to an axisymmetric indenter of geometry defined by the function *f(r)* by using a known pressure distribution due to another indenter $P^*(r)$ (e.g. a flat punch cylinder) [1], [2]

$$F = \frac{2}{\pi}\int_0^a P^*(r)\left[\delta - f(r)\right] r\,dr \tag{S1}$$

In addition, the indentation in terms of the contact area can be obtained by imposing that [1], [2]

$$\frac{\partial F(\delta, a)}{\partial a} = 0 \tag{S2}$$

which is reminiscent of the zero-pressure condition outside the contact area. The problem then reduces to finding the pressure distribution of a single indenter, which is substituted into Eq. (S1) and performing the integration. The simplest indenter to consider mathematically is the flat punch cylinder. To obtain the pressure distribution for a finite-thickness film, one can develop the Green's function for the problem in a series of expansions and solve the corresponding set of recursive integral equations. Following this procedure, for the flat cylindrical punch it has been shown that [1]

$$P^*(r) = \frac{E}{\pi(1-\nu^2)}\left(\frac{\delta}{a}\right)\left[\bar{P}_1\left(\frac{a}{h}\right)\frac{1}{\sqrt{1-\left(\frac{r}{a}\right)^2}} + \bar{P}_2\left(\frac{a}{h}\right)\frac{\left(\frac{r}{a}\right)^2}{\sqrt{1-\left(\frac{r}{a}\right)^2}}\right] \tag{S3}$$

where

$$\bar{P}_1\left(\frac{a}{h}\right) = 1 - \frac{2\alpha_0}{\pi}\left(\frac{a}{h}\right) + \left(\frac{2\alpha_0}{\pi}\right)^2\left(\frac{a}{h}\right)^2 - \frac{8}{\pi}\left(\frac{\alpha_0^{\,3}}{\pi^2} - \frac{\beta_0}{3}\right)\left(\frac{a}{h}\right)^3 + \frac{16}{\pi^4}\alpha_0^4\left(\frac{a}{h}\right)^4 \tag{S4}$$

$$\bar{P}_2\left(\frac{a}{h}\right) = \left(\frac{a}{h}\right)^2\left[-\frac{8\beta_0}{\pi}\left(\frac{a}{h}\right) + \frac{16\alpha_0\beta_0}{\pi^2}\left(\frac{a}{h}\right)^2\right] \tag{S5}$$

with [1]

$$\alpha_0(\nu) = -\frac{1.2876 - 1.4678\nu + 1.3442\nu^2}{1-\nu}, \qquad \beta_0(\nu) = \frac{0.6387 - 1.0277\nu + 1.5164\nu^2}{1-\nu} \tag{S6}$$

Equation (S6) is a phenomenological function obtained by solving numerically some of the integrals in the derivation. For the case of a spherical tip, one has

$$f(r) = R - \sqrt{R^2 - r^2} \tag{S7}$$

By substituting Eqs. (S3)-(S6) into Eq. (S1) and using the integrals

$$\int_0^a \frac{r}{\sqrt{a^2-r^2}}dr = a$$

$$\int_0^a \frac{r^3}{\sqrt{a^2-r^2}}dr = \frac{2}{3}a^3$$

$$\int_0^a \frac{r}{\sqrt{a^2-r^2}}\left(R-\sqrt{R^2-r^2}\right)dr = aR+\frac{1}{2}\left(-aR+\left(a^2-R^2\right)\text{arctanh}\left(\frac{a}{R}\right)\right)$$

$$\int_0^a \frac{r^4}{\sqrt{a^2-r^2}}\left(R-\sqrt{R^2-r^2}\right)dr = \frac{7}{24}a^3R+\frac{1}{8}aR^3+\frac{1}{8}\left(a^2-R^2\right)\left(3a^2+R^2\right)\text{arctanh}\left(\frac{a}{R}\right)$$

(S8)

one finally obtains

$$F=\frac{2Eh^3}{\left(1-\nu^2\right)R}\left(\frac{a}{h}\right)\left\{\left[\overline{P}_1\left(\frac{a}{h}\right)+\frac{2}{3}\overline{P}_2\left(\frac{a}{h}\right)\right]\left(\frac{\delta R}{h^2}\right)\right.$$
$$\left.-\left(\frac{Ra}{h^2}\right)\overline{P}_1\left(\frac{a}{h}\right)\frac{1}{2}\left(\frac{R}{a}+\left(1-\left(\frac{R}{a}\right)^2\right)\text{arctanh}\left(\frac{a}{R}\right)\right)-\left(\frac{Ra}{h^2}\right)\overline{P}_2\left(\frac{a}{h}\right)\frac{1}{8}\left(\frac{7}{3}\left(\frac{R}{a}\right)+\left(\frac{R}{a}\right)^3+\left(1-\left(\frac{R}{a}\right)^2\right)\left(3+\left(\frac{R}{a}\right)^2\right)\text{arctanh}\left(\frac{a}{R}\right)\right)\right\} \quad \text{(S9)}$$

By imposing the boundary condition in Eq. (S2), the relationship between the indentation and the contact area is given by

$$\frac{R\delta}{h^2}=\left(\frac{a}{h}\right)^2\left(\frac{R}{a}\right)\left\{\frac{\overline{P}_{1d}\left(\frac{a}{h}\right)\frac{1}{2}\left[\frac{R}{a}+\left(1-\left(\frac{R}{a}\right)^2\right)\text{arctanh}\left(\frac{a}{R}\right)\right]+\overline{P}_1\left(\frac{a}{h}\right)\text{arctanh}\left(\frac{a}{R}\right)}{\overline{P}_{1d}\left(\frac{a}{h}\right)+\overline{P}_1\left(\frac{a}{h}\right)+\frac{2}{3}\overline{P}_{2d}\left(\frac{a}{h}\right)+2\overline{P}_2\left(\frac{a}{h}\right)}+\right.$$
$$\left.\frac{\overline{P}_{2d}\left(\frac{a}{h}\right)\frac{1}{8}\left[\frac{7}{3}\frac{R}{a}+\left(\frac{R}{a}\right)^3+\left(1-\left(\frac{R}{a}\right)^2\right)\left(3+\left(\frac{R}{a}\right)^2\right)\text{arctanh}\left(\frac{a}{R}\right)\right]}{\overline{P}_{1d}\left(\frac{a}{h}\right)+\overline{P}_1\left(\frac{a}{h}\right)+\frac{2}{3}\overline{P}_{2d}\left(\frac{a}{h}\right)+2\overline{P}_2\left(\frac{a}{h}\right)}+\frac{\overline{P}_2\left(\frac{a}{h}\right)\frac{1}{2}\left[\frac{R}{a}+\left(3-\left(\frac{R}{a}\right)^2\right)\text{arctanh}\left(\frac{a}{R}\right)\right]}{\overline{P}_{1d}\left(\frac{a}{h}\right)+\overline{P}_1\left(\frac{a}{h}\right)+\frac{2}{3}\overline{P}_{2d}\left(\frac{a}{h}\right)+2\overline{P}_2\left(\frac{a}{h}\right)}\right\} \quad \text{(S10)}$$

where

$$\overline{P}_{1d}\left(\frac{a}{h}\right)=\left(\frac{a}{h}\right)\left[-\frac{2\alpha_0}{\pi}+\left(\frac{2\alpha_0}{\pi}\right)^2 2\left(\frac{a}{h}\right)-\frac{8}{\pi}\left(\frac{\alpha_0}{\pi^2}-\frac{\beta_0}{3}\right)3\left(\frac{a}{h}\right)^2+\frac{16}{\pi^4}\alpha_0^4 4\left(\frac{a}{h}\right)^3\right] \quad \text{(S11)}$$
$$\overline{P}_{2d}\left(\frac{a}{h}\right)=\left(\frac{a}{h}\right)^3\left[-\frac{8\beta_0}{\pi}+\frac{16\alpha_0\beta_0}{\pi^2}2\left(\frac{a}{h}\right)\right]$$

Performing a series expansion in Eq. (S10) in terms of *a/h* up to O(4), one obtains

$$\delta \approx a\text{arctanh}\left(\frac{a}{R}\right)+\left(-aR+\left(R^2+a^2\right)\text{arctanh}\left(\frac{a}{R}\right)\right)\left[\frac{\alpha_0}{\pi}\frac{1}{h}+8\frac{\alpha_0\beta_0}{3\pi^2}a^3\left(\frac{1}{h}\right)^4\right]$$
$$+\frac{\beta_0}{3\pi}\left(\left(3R^4+6a^2R^2+7a^4\right)\text{arctanh}\left(\frac{a}{R}\right)-aR\left(3R^2+7a^2\right)\right)\left(\frac{1}{h}\right)^3 \quad \text{(S12)}$$

Moreover, by substituting Eq. (S12) into Eq. (S10) and performing the series expansion of the resulting expression up to O(4) in *a/h* one obtains

$$F \approx \frac{E}{\left(1-\nu^2\right)}\left(\left(R^2+a^2\right)\operatorname{arctanh}\left(\frac{a}{R}\right)-aR\right)\left[1+\frac{8\beta_0}{3\pi}\left(\frac{a}{h}\right)^3\right] \tag{S13}$$

For $\nu$=0.5, one has $\alpha_0$=−1.7795 and $\beta_0$=1.0079 (see Eq. (S6)). By substituting these values into Eqs. (S12) and (S13), one obtains Eq. (B9). Equations (S12) and (S13) can be used to obtain the version of Eq. (B9) for other values of the Poisson's ratio.

## S7. Analysis of non-linear effects

For the case of a spherical indenter, Long et al. [3] performed finite element numerical calculations for a neo-Hookean elastic thin film model and proposed the following phenomenological expression to describe the force indentation curves for the case of frictionless non-adhesive contact and bottom-bonded condition:

$$F_{Long} = \frac{4}{3}\frac{E}{\left(1-0.5^2\right)}\sqrt{R}\delta^{3/2}\left(1+1.15\left(\frac{R\delta}{h^2}\right)^{1/3}+\alpha\left(R/h\right)\left(\frac{R\delta}{h^2}\right)+\beta\left(R/h\right)\left(\frac{R\delta}{h^2}\right)^2\right) \qquad \frac{\delta}{h}\leq \min\left(0.6, R/h\right) \qquad \text{(S14)}$$

$$\alpha\left(\frac{R}{h}\right)=10.05-0.63\sqrt{\frac{h}{R}}\left(3.1+\frac{h^2}{R^2}\right); \quad \beta\left(\frac{R}{h}\right)=4.8-4.23\frac{h^2}{R^2} \qquad 0.3\leq\frac{R}{h}\leq 12.7$$

This model includes finite indentation and finite thickness effects, as well as non-linear elastic behaviour. Fig. S4a shows the comparison of this non-linear model with the linear model proposed in the present work.

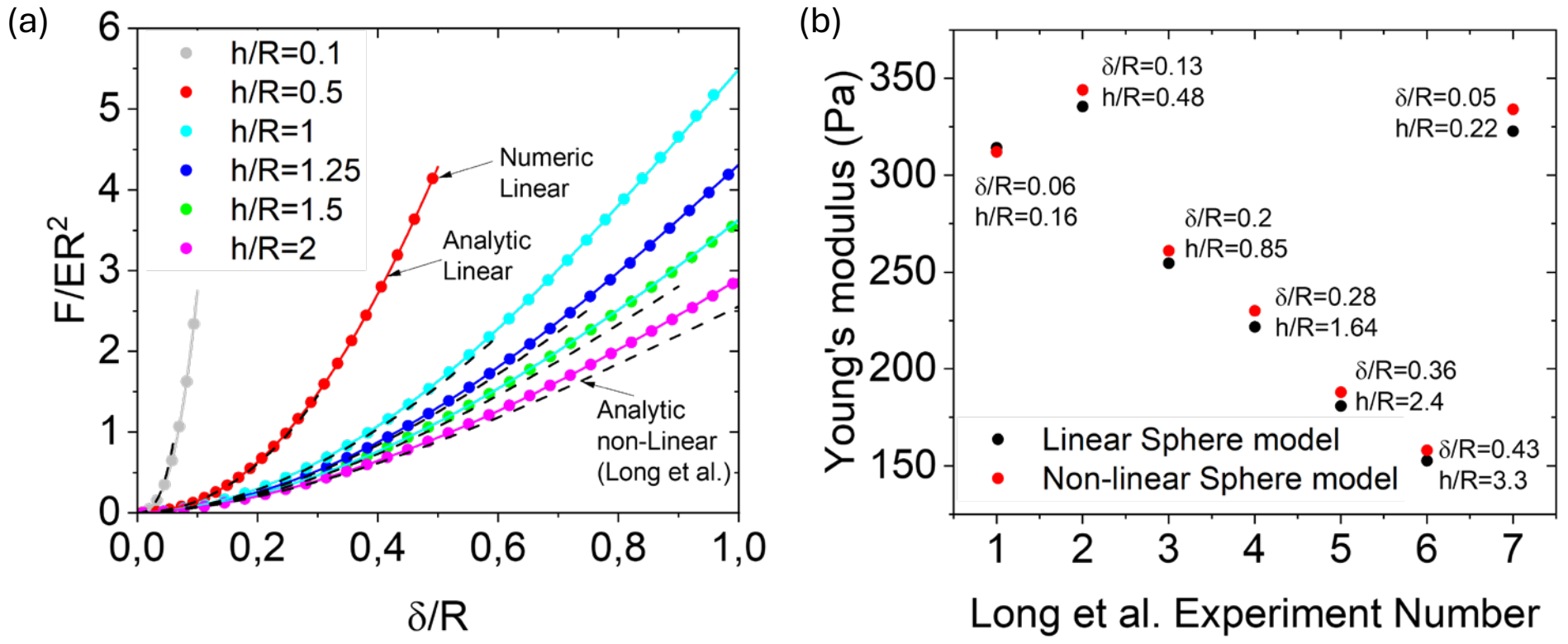

**Figure S4.** (a) Comparison of the normalised force indentation curves for a Poisson's ratio, $\nu$=0.5, predicted by the non-linear elastic phenomenological model of Long et al. [3] (black dashed lines), the linear elastic model presented in the present work (colour lines) and the numerical calculations for a linear elastic model (symbols). In all cases, frictionless contact and bottom-bonded conditions are assumed. (b) Re-analysis of the data reported in Ref. [3] by using the sphere model presented here.

In its range of validity, Long's non-linear model predictions are quite similar to the results of our linear model, showing at most a relative error of 13%. Note that for indentations $d/R$<0.4 both models predict similar behaviours. Figure S4b shows the extracted Young's modulus from the experiments reported by Long et al. [3] by using Long's non-linear elastic model and the sphere elastic model presented in the present work. For this experimental data, the maximum relative error observed is just 3%. The errors are relatively low because the experiments have explored moderate indentation values, $\delta/R$<0.48, where both models are practically identical (see Fig. S4a).

**Bibliography**

[1] E. K. Dimitriadis, F. Horkay, J. Maresca, B. Kachar, and R. S. Chadwick, "Determination of elastic moduli of thin layers of soft material using the atomic force microscope," *Biophys J*, vol. 82, no. 5, pp. 2798–2810, 2002, doi: 10.1016/S0006-3495(02)75620-8.

[2] P. D. Garcia and R. Garcia, "Determination of the Elastic Moduli of a Single Cell Cultured on a Rigid Support by Force Microscopy," *Biophys J*, vol. 114, no. 12, pp. 2923–2932, Jun. 2018, doi: 10.1016/j.bpj.2018.05.012.

[3] R. Long, M. S. Hall, M. Wu, and C. Y. Hui, "Effects of gel thickness on microscopic indentation measurements of gel modulus," *Biophys J*, vol. 101, no. 3, pp. 643–650, Aug. 2011, doi: 10.1016/j.bpj.2011.06.049.